\documentclass[useAMS,usenatbib,usegraphicx]{mn2e}

\usepackage{aas_macros}

\title[Disc formation in turbulent, magnetised cores]{Turbulence-induced disc formation in strongly magnetised cloud cores}
  \author[D. Seifried et al.]
  {D.~Seifried,$^{1,2,3}$\thanks{dseifried@hs.uni-hamburg.de} R.~Banerjee,$^{1}$ R.~E.~Pudritz,$^{3,4}$ R.~S.~Klessen$^2$ \\
  $^1$Hamburger Sternwarte, Universit\"at Hamburg, Gojenbergsweg 112, 21029 Hamburg, Germany\\
  $^2$Universit\"at Heidelberg, Zentrum für Astronomie, Institut f\"ur Theoretische Astrophysik, Albert-Ueberle-Str. 2, 69120 Heidelberg, Germany \\
  $^3$Department of Physics $\&$ Astronomy, McMaster University, Hamilton ON L8S 4M1, Canada\\
  $^4$Origins Institute, McMaster University, ABB 241, Hamilton ON L8S 4M1, Canada}

\date{Released 2012}

\pagerange{\pageref{firstpage}--\pageref{lastpage}} \pubyear{2013}

\begin{document}

\label{firstpage}

\maketitle

\begin{abstract}
We present collapse simulations of strongly magnetised, turbulent molecular cloud cores with masses ranging from 2.6 to 1000 M$_{\sun}$ in order to study the influence of the initial conditions on the turbulence-induced disc formation mechanism proposed recently by~\citet{Seifried12}. We find that Keplerian discs are formed in all cases independently of the core mass, the strength of turbulence, or the presence of global rotation. The discs appear within a few kyr after the formation of the protostar, are 50 -- 150 AU in size, and have masses between 0.05 and a few 0.1 M$_{\sun}$. During the formation of the discs the mass-to-flux ratio stays well below the critical value of 10 for Keplerian disc formation. Hence, flux-loss alone cannot explain the formation of Keplerian discs. The formation of rotationally supported discs at such early phases is rather due to the disordered magnetic field structure and due to turbulent motions in the surroundings of the discs, two effects lowering the classical magnetic braking efficiency. Binary systems occurring in the discs are mainly formed via the disc capturing mechanism rather than via disc fragmentation, which is largely suppressed by the presence of magnetic fields.
\end{abstract}


\section{Introduction}

The formation of protostellar discs under the influence of magnetic fields has received a great deal of attention for about one decade~\citep[e.g.][]{Allen03,Matsumoto04,Machida05,Banerjee06,Banerjee07,Price07,Hennebelle08,Hennebelle09,Duffin09,Commercon10,Peters11,Seifried11}. In these works a highly idealised numerical setup usually consisting of an overall core rotation and a magnetic field parallel to the rotation axis was used. One common finding of these simulations is that for an initial setup with a magnetic field strength comparable to observations, i.e. a mass-to-flux ratio smaller than 5 -- 10~\citep[e.g.][]{Falgarone08,Girart09,Beuther10}, no rotationally supported discs were found. This result is also called the ''magnetic braking catastrophe'' owing to the fact that magnetic braking is responsible for the removal of the angular momentum which would be necessary to form the discs. However, these results seem to be in contrast to recent high-resolution observations revealing the existence of Keplerian discs around Class 0 protostellar objects~\citep[e.g.][]{Tobin12}.

Trying to circumvent this ''magnetic braking catastrophe'' by the inclusion of non-ideal MHD effects seems to fail. For instance the inclusion of ambipolar diffusion~\citep[e.g.][]{Mellon09,Duffin09} does not result in the formation of Keplerian discs in the earliest phase of protostellar evolution. Furthermore, Ohmic dissipation does also not allow for large-scale ($\sim$ 100 AU) but only for very small ($\sim$ 10 solar radii) rotationally supported structures~\citep[e.g.][]{Dapp11}, unless one assumes an unusually high resistivity~\citep{Krasnopolsky10}. Even the combined effects of ambipolar diffusion and Ohmic dissipation cannot reduce the magnetic braking efficiency significantly~\citep{Li11}. Recently, a third non-ideal MHD effect namely the Hall effect was shown to enable the formation of Keplerian discs~\citep{Krasnopolsky11}. However, this seems to require a Hall coefficient about one order of magnitude larger than expected under realistic conditions. Furthermore, the formation of a disc in these simulations is not due to a reduced magnetic braking efficiency but due to the spin-up of the gas in the midplane by the Hall effect as the authors clearly demonstrated by the occurrence of a disc spinning in the opposite direction than the surrounding protostellar core.

As mentioned before, all the simulations use a highly idealised setup of a uniformly rotating core and a magnetic field parallel to the rotation axis. Only recently have the effects of deviating from such a highly idealised setup been included. \citet{Hennebelle09},~\citet{Ciardi10} and~\citet{Joos12} investigated how an inclination between the overall magnetic field and the core rotation affects the formation of Keplerian discs. The authors found that even for a moderate inclination the formation of rotationally supported discs is possible. However, as~\citet{Krumholz13} pointed out recently, for a realistic distribution of magnetic field strengths and misalignment angles, this would lead to a fraction of Keplerian discs between 10 -- 50\% only. In an alternative approach, the inclusion of turbulent motions in the initial conditions and their effect on the formation of rotationally supported discs was studied~\citep{Seifried12,Santos12b,Santos12}, again showing the possibility of forming Keplerian discs for strong magnetic fields. \citet{Santos12b,Santos12} attribute the formation to turbulent reconnection occurring in the region of the disc, which in turn reduces the magnetic flux and thus the magnetic braking efficiency. In contrast, in~\citet{Seifried12} we suggest that the build-up of Keplerian discs is due to the turbulent motions in the surroundings of the discs rather than due to magnetic flux loss. These motions distort the magnetic field and hamper the build-up of a toroidal magnetic field component responsible for angular momentum extraction, but simultaneously provide a sufficient amount of angular momentum necessary for disc formation.

In this work we investigate this -- as we call it -- turbulence-induced formation mechanism in more detail and for a wider range of initial conditions. In~\citet{Seifried12} we tested this mechanism only for a massive molecular cloud of 100 M$_{\sun}$ and supersonic turbulence. Since we consider the turbulent motions to be the main driver of the formation of Keplerian discs, it is particularly interesting to test this mechanism for the weaker case of subsonic turbulence. As subsonic turbulence is expected to dominate in low-mass molecular cloud cores, we extend our analysis also to the low-mass range. Moreover, we study if or to what extent the turbulence-induced formation mechanism depends on the presence of an overall uniform rotation. We show that in the presence of turbulence in essentially any case Keplerian discs are formed by the aforementioned mechanism.

The paper is organised as follows. In Section~\ref{sec:techniques} we present the details of the initial conditions of the simulations and briefly describe the numerical techniques. The result of the simulations are presented are in Section~\ref{sec:results}. First, we present the details of the Keplerian discs formed and analyse possible reasons for their formation. Next, we consider the time evolution of global disc properties before studying the fragmentation properties of the discs in more detail. In Section~\ref{sec:discussion} the results are discussed in a broader context and are compared to related numerical and observational work before we summarise our main findings in Section~\ref{sec:conclusion}.

\section{Initial conditions}
\label{sec:techniques}

We now describe the setup of the different simulations. We performed several simulations with core masses ranging from to 2.6 M$_{\sun}$ to 1000 M$_{\sun}$ (see Table~\ref{tab:models} for an overview over the different models and their parameters). All cores are embedded in a cubic box with a length of approximately three times the cores diameter to avoid corruption of the results by boundary effects. The ambient medium has a density a factor of 100 lower than the density at the edge of the cores, hence its dynamical influence can be considered negligible. The runs 2.6-NoRot-M2 and 2.6-Rot-M2 have a core mass of 2.6 M$_{\sun}$ and a gas temperature of 15 K\footnote{The ambient gas has a 100 times higher gas temperature to assure pressure equilibrium at the edge of the core.}. Here we note that the first part in the name of each run denotes the mass-to-flux ratio, the second whether a uniform rotation is present or not, and the third the (approximate) core mass in M$_{\sun}$. The remaining runs with higher core masses have a initial temperature of 20 K, thus somewhat higher than in the low-mass cores~\citep[see e.g.][]{Ragan12,Launhardt13}.

In all runs we set up a Kolmogorov-type turbulence field so that 3D-turbulent rms Mach number ($M_\rmn{rms}$) ranges from subsonic turbulence ($M_\rmn{rms} \sim 0.74$) in the low-mass runs to strongly supersonic turbulence in the high-mass cases ($M_\rmn{rms} \sim 2.5 - 5.4$). These values represent the typical level of turbulence observed in molecular cloud cores~\citep[e.g.][but see also the review of~\citealt{Ward07} and references therein]{Caselli95,Andre07} and result in a ratio of the turbulent energy to gravitational energy $\beta_\rmn{turb}$ of about 0.08 to 0.12 (see Table~\ref{tab:models}). Here the gravitational energy is determined from the analytical expression for the given density profile (see further down)\footnote{We note that in our previous work a value for $\beta$ of 0.04 was given. This value resulted from the calculation of the gravitational energy from the simulation data not taking into account a systematic offset in the numerical gravitational potential compared to the analytical value, which results in the somewhat lower value for $\beta$.}. For the runs with 100 M$_{\sun}$ we performed several simulations in order to test to what extent the results depend on the random realisations of the turbulence field (seed A,B, and C). We found that the results do not change qualitatively if the turbulence field is varied~\citep[but see][for a detailed discussion]{Seifried12}. In some of the runs in addition we superimpose a rigid rotation of the core around the $z$-axis on the turbulent velocity field. For the sake of simplicity the rotation frequency in these runs is chosen such that the rotation energy equals the turbulent kinetic energy, i.e. $\beta_\rmn{rot} = \beta_\rmn{turb}$. All cores are threaded by a magnetic field parallel to the $z$-axis. The strength of the field declines with increasing cylindrical radius $R$ as
\begin{equation}
 B_z \propto \sqrt{\rho(R)_\rmn{midplane}} \, ,
\end{equation}
where $\rho(R)_\rmn{midplane}$ is the density in the midplane at the radius $R$. This guarantees that $\beta_\rmn{plasma} = P/(B^2/8\pi)$ is constant in the equatorial plane, where $P$ is the thermal pressure. Furthermore, in order to guarantee $\nabla \bmath{B} = 0$, $B_z$ does not vary along  the $z$-axis. The strength of the magnetic field is chosen such that the (normalised) mass-to-flux ratio $\mu$ of the cores is 2.6. We emphasise that throughout the paper the mass-to-flux ratio is given in units of the critical mass-to-flux ratio $\mu_\rmn{crit} = 0.13/\sqrt{G}$~\citep{Mouschovias76}, where $G$ is the gravitational constant.
\begin{table*}
 \caption{Initial conditions of the performed simulations showing the mass, radius, normalised mass-to-flux ratio $\mu$, whether uniform rotation is present or not, the angular frequency, the turbulent energy content normalised to the gravitational energy, the seed of the random realisation of the turbulence field, the power spectrum index p of the turbulence spectrum, the rms Mach number, and simulated physical times of the protostellar discs.}
 \label{tab:models}
 \begin{tabular}{@{}lccccccccccc}
  \hline
  Run & $m_\rmn{core}$ & r$_\rmn{core}$ & $\mu$  & rotation & $\Omega$ & $\beta_{\rmn{turb}}$ & turbulence & p & $M_\rmn{rms}$ & t$_\rmn{sim}$ \\
      &  [M$_{\sun}$]  & [pc] & & & [10$^{-13}$ s$^{-1}$] & & seed & & & [kyr] \\
  \hline
  2.6-NoRot-M2 & 2.6 & 0.0485 & 2.6 & No & 0 & 0.087 & A & 5/3 & 0.74 & 15 \\
  2.6-Rot-M2 & 2.6 & 0.0485 & 2.6 & Yes & 2.20 & 0.087 & A & 5/3 & 0.74 & 15 \\
  2.6-NoRot-M100 & 100 & 0.125 & 2.6 & No & 0 & 0.084 & A & 5/3 & 2.5 & 15 \\
  2.6-Rot-M100 & 100 & 0.125 & 2.6 & Yes & 3.16 & 0.084 & A & 5/3 & 2.5 & 15 \\
  2.6-Rot-M100-B & 100 & 0.125 & 2.6 & Yes & 3.16 & 0.084 & B & 5/3 & 2.5 & 15 \\
  2.6-Rot-M100-C & 100 & 0.125 & 2.6 & Yes & 3.16 & 0.084 & C & 5/3 & 2.5 & 15 \\
  2.6-Rot-M100-p2 & 100 & 0.125 & 2.6 & Yes & 3.16 & 0.084 & A & 2 & 2.5 & 15 \\
  2.6-NoRot-M300 & 300 & 0.125 & 2.6 & No & 0 & 0.12 & A & 5/3 & 5.0 & 10 \\
  2.6-Rot-M1000 & 1000 & 0.375 & 2.6 & Yes & 1.90 & 0.081 & A & 5/3 & 5.4 & 10 \\
  \hline
 \end{tabular}
\end{table*}

The density profile varies between the different runs. In the runs 2.6-NoRot-M2 and 2.6-Rot-M2 the density profile follows that of a Bonnor-Ebert (BE) sphere up to a radius of $6.9 r_0$ ($6.49 r_0$ is the critical radius for a BE sphere). The physical radius of the core is 10\,000 AU, and the density of the BE-profile is scaled up in order to contain 2.6 M$_{\sun}$ in the core. Therefore, the core is slightly gravitational unstable, which is reasonable since additional effects like rotation, turbulence and magnetic fields can counteract gravity. The runs with 100 M$_{\sun}$ and 300 M$_{\sun}$ have a diameter of 0.125 pc and are highly gravitationally unstable (56 and 290 Jeans masses, respectively). The density of the cores declines outwards as\footnote{To avoid unphysically high densities in the interior of the core, we cut off the $r^{-1.5}$ profile at a radius of 0.0125 pc. Within this radius the density distribution follows a parabola $\rho(r) \propto [1 - (r/r_0)^2]$.}
\begin{equation}
 \rho(r) \propto r^{-1.5} \, .
 \label{eq:profile}
\end{equation}
Since the cores are highly gravitationally unstable, this explains the higher turbulent Mach numbers compared to the runs with 2.6 M$_{\sun}$ despite comparable values of $\beta_\rmn{turb}$. Run 2.6-Rot-M1000 has a mass of 1000 M$_{\sun}$ corresponding to about 340 Jeans masses contained in a core with a radius of 0.375 pc having qualitatively the same density profile as described in Eq.~\ref{eq:profile}.

The cooling routine applied in all runs takes into account dust cooling, molecular line cooling and the effects of optically thick gas~\citep{Banerjee06}. We introduce sink particles above a density threshold of $\rho_{\rmn{crit}} = 1.14 \cdot 10^{-10}$ g cm$^{-3}$~\citep[see][for details]{Federrath10}. The maximum spatial resolution is set to 1.2 AU in all simulations. The refinement criterion used guarantees that the Jeans length is resolved everywhere with at least 8 grid cells.

\section{Results}
\label{sec:results}

In what follows we focus on the effects of varying core masses and initial rotation on the formation of early-type discs. For this reason we present details of the runs 2.6-NoRot-M2, 2.6-Rot-M2, 2.6-NoRot-M100, 2.6-Rot-M100, 2.6-NoRot-M300 and 2.6-Rot-M1000 only. The remaining runs with core masses of 100 M$_{\sun}$ listed in Table~\ref{tab:models} were discussed in detail in~\citet{Seifried12}. We again emphasise that we did not find any qualitative differences for different realisations of the turbulence field. We therefore believe that also in the runs with lower and higher masses a change in the turbulence field would not significantly alter the results~\citep[see also][]{Joos13}.

\subsection{Disc formation}
\label{sec:discs}

\subsubsection{Column density structure}

Our first step is to compute the column density structure of the discs formed in the simulations and to analyse how they evolve over time. We note that the column density is computed for the discs for the face on view. For the calculation of the disc's angular momentum we only take into account gas with a density larger than $5 \cdot 10^{-13}$ g cm$^{-3}$. Next, the angular momentum is calculated with respect to the centre-of-mass of the disc. The threshold density of $5 \cdot 10^{-13}$ g cm$^{-3}$ is physically motivated by the fact that gas gets optically thick in this range (see also the discussion in Section~\ref{sec:compare}). We found that when using this threshold the radius of the disc also reasonably well agrees with the rather sharp drop-off in the column density marking the outer edge of the disc (see figures below). Furthermore, making use of a density threshold has the advantage that gas from the outflow cavity above and below the disc, which usually has significantly lower densities, is excluded.

\begin{figure}
 \centering
 \includegraphics[width=82mm]{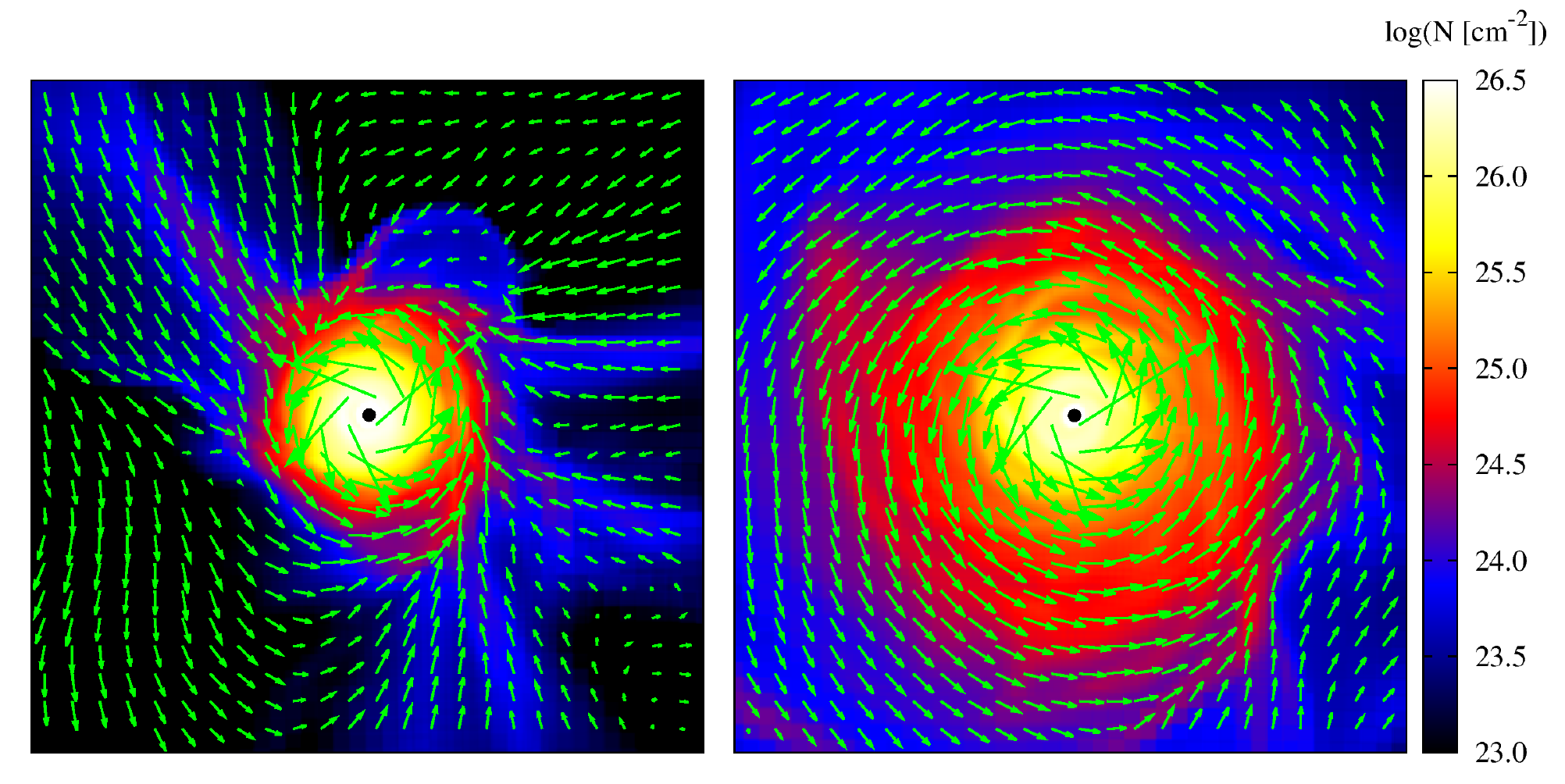}
 \includegraphics[width=82mm]{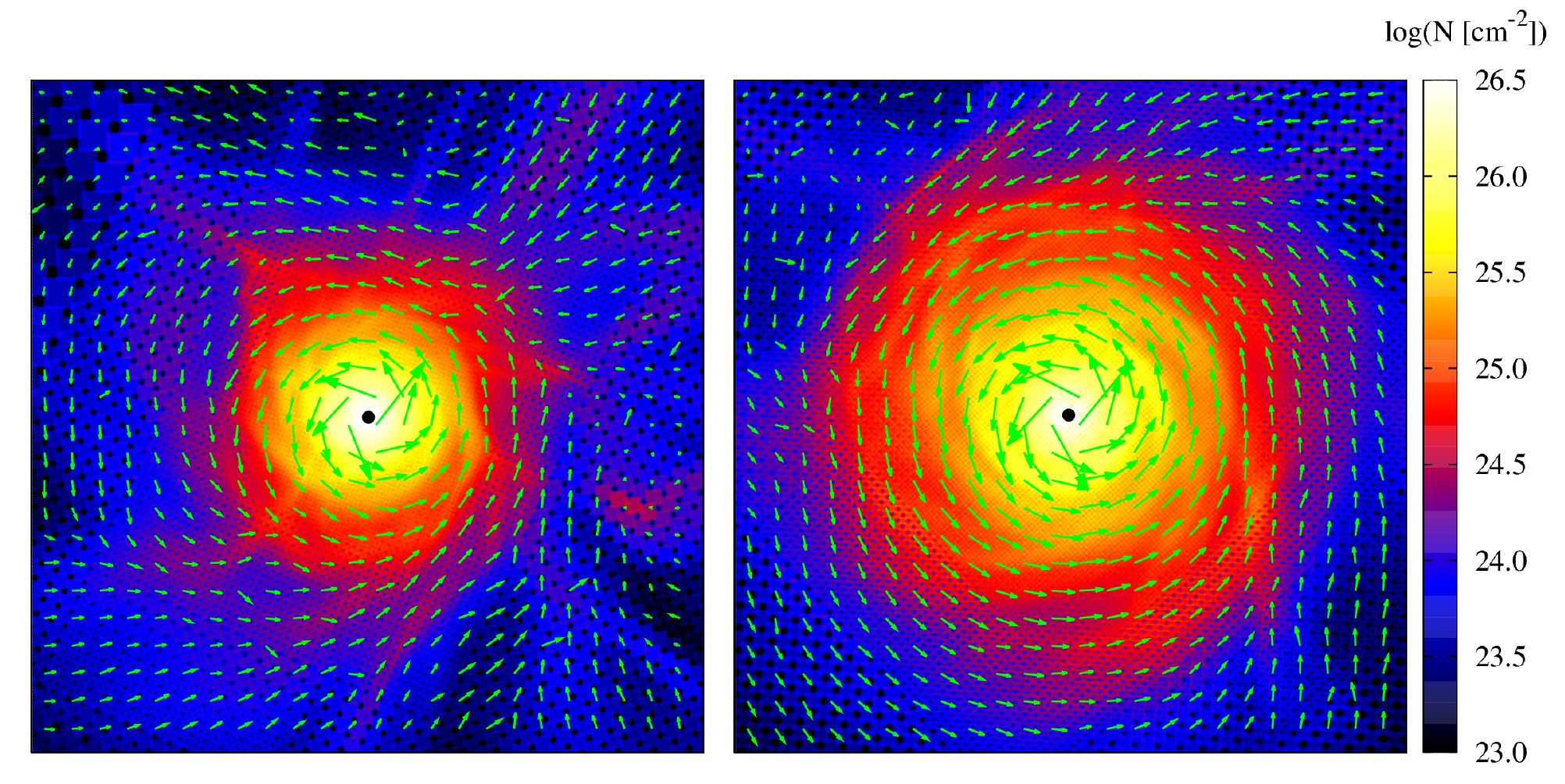}
\caption{Column density in logarithmic scaling for the discs in run 2.6-NoRot-M2 (top) and 2.6-Rot-M2 (bottom) for 7.5 kyr (left) and 15 kyr (right) after the formation of the first sink particle. The viewing direction was adjusted in order to see the discs face on. The green arrows show the mass-weighted mean velocity and the black dots the projected positions of the sink particles. The figures are 200 AU in size.}
 \label{fig:diskM2}
\end{figure}

We plot the column density structure of the six runs mentioned in the beginning of this section for two different times, namely half way from the formation of the first sink particle towards the end of the simulation (7.5 and 5 kyr, respectively) and directly at the end of each simulation (15 and 10 kyr, respectively). In addition, we also show the line-of-sight averaged velocity in the discs. First we show the results for the discs of the runs with 2.6 M$_{\sun}$ cores in Fig.~\ref{fig:diskM2}. The viewing direction was adjusted to the direction of the angular momentum vector such that the discs are seen face on. Due to the rather weak turbulent motions and the marginally gravitationally unstable configuration of the cores, in both runs no fragmentation has occurred and only one sink particle with an associated disc has formed so far.
Furthermore, it can be seen that in both runs a rotationally supported structure has built up after 7.5 kyr and that the disc in run 2.6-Rot-M2 is somewhat larger that that in run 2.6-NoRot-M2. 

\begin{figure}
 \centering
 \includegraphics[width=82mm]{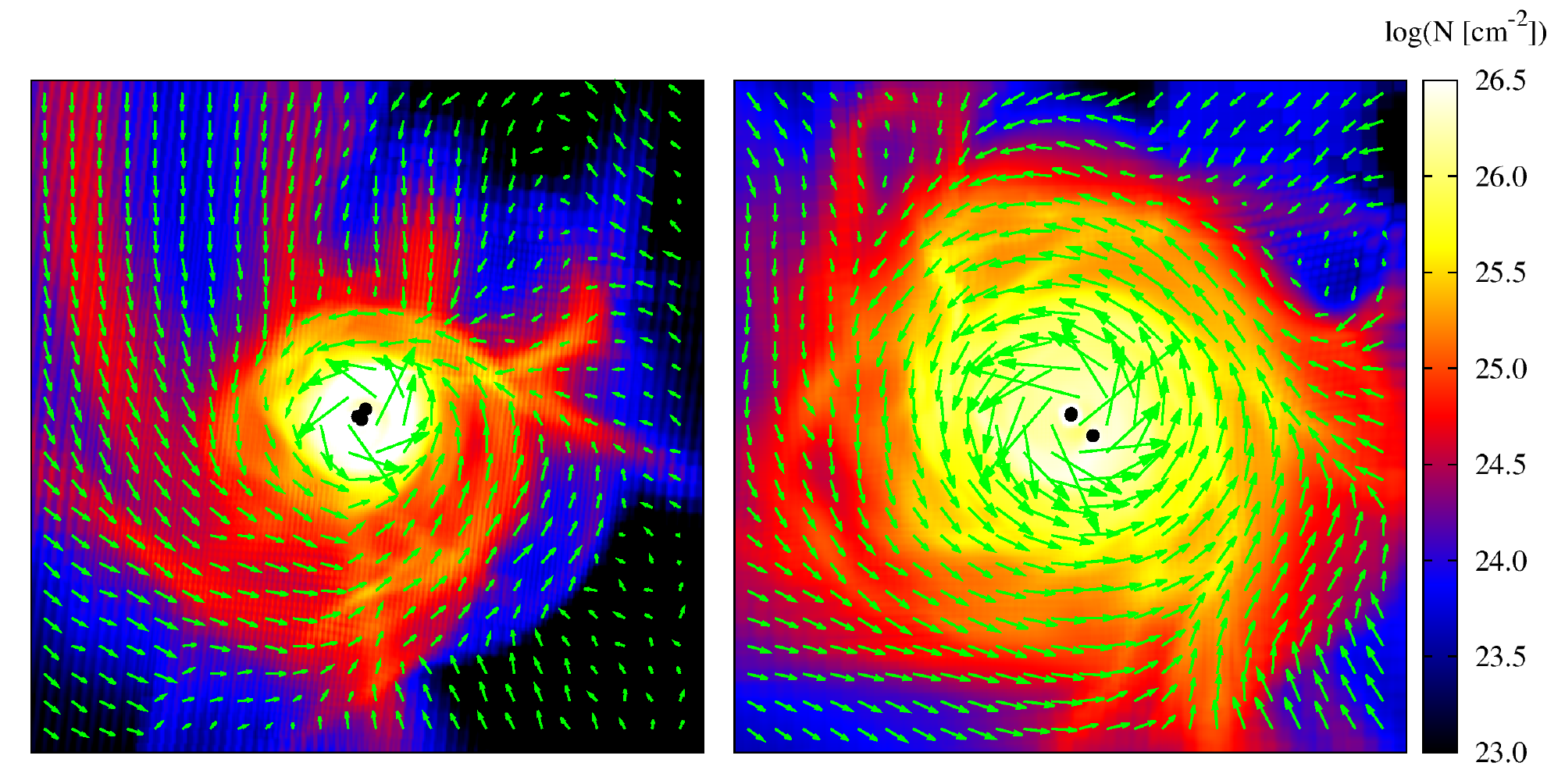}
 \includegraphics[width=82mm]{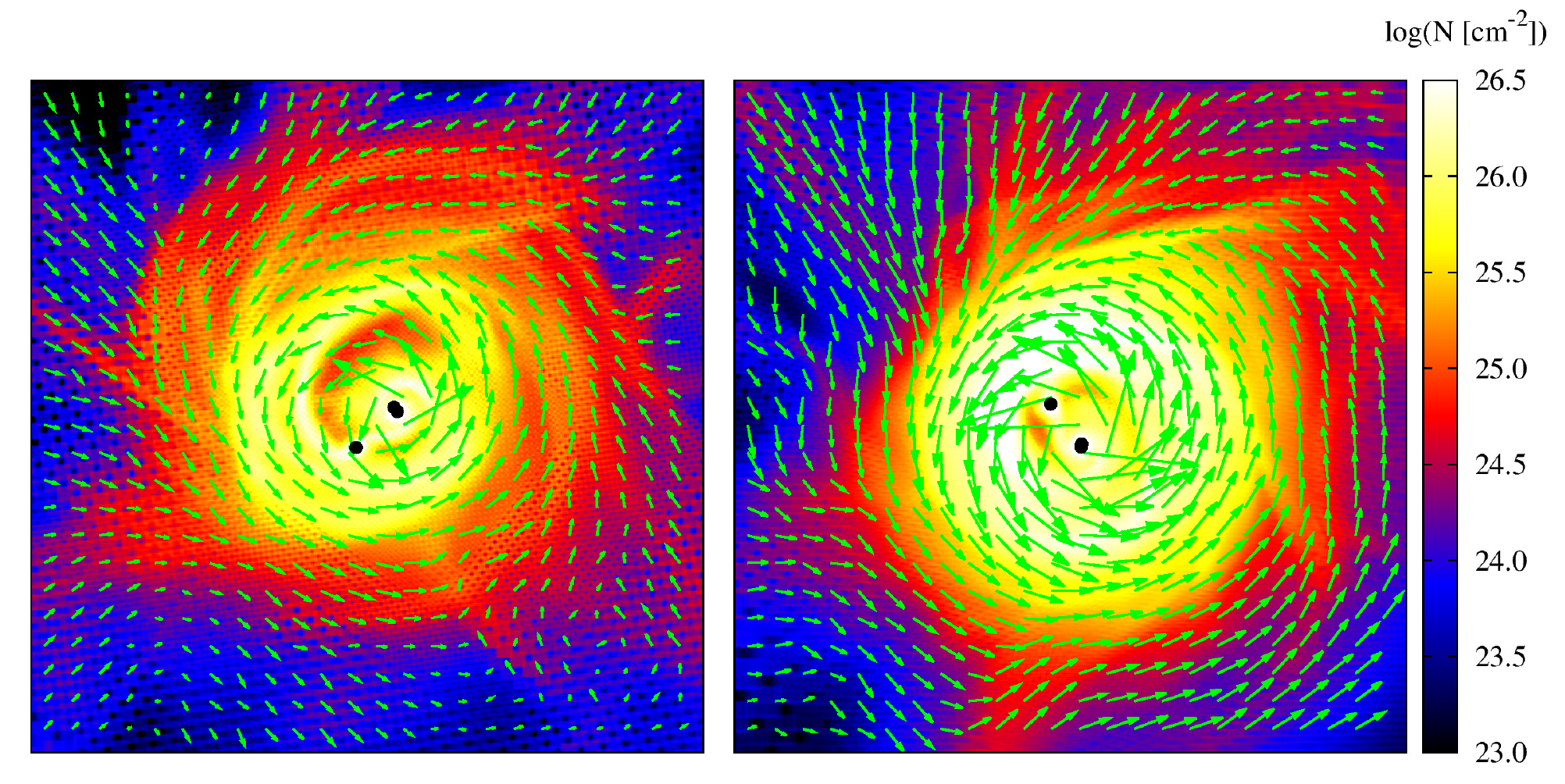}
\caption{Same as in Fig.~\ref{fig:diskM2} but for run 2.6-NoRot-M100 (top) and run 2.6-Rot-M100 (bottom).}
 \label{fig:diskM100}
\end{figure}

In Fig.~\ref{fig:diskM100} we plot the structure of the discs in run 2.6-NoRot-M100 and run 2.6-Rot-M100. Due to the supersonic turbulent motions and the highly gravitationally unstable configuration, the cores in both runs have fragmented and formed 5 sinks by the end of each run. However, only one disc has formed so far in both runs (see Section~\ref{sec:fragmentation} for the details of the fragmentation process). In run 2.6-Rot-M100 the three most massive sinks are all contained within one massive disc. The two remaining sink particles are a few 1000 AU off from the centre of this disc and do both not contain any associated disc, i.e. no gas with densities above $5 \cdot 10^{-13}$ g cm$^{-3}$ has been found in their surroundings. We attribute this to the fact that both sink particles have been ejected from the massive disc before and are thus moving with a high relative speed compared to the surrounding medium, which prevents the build-up of discs. Similar holds for run 2.6-NoRot-M100 where one sink has been ejected and one sink has been formed shortly before the end of the simulation so that nothing can be said about an associated disc, yet. The remaining three sink particles are again grouped in one disc.
Like in the runs with 2.6 M$_{\sun}$, in both runs the discs show significant rotationally support already at 7.5 kyr. Furthermore, it can be inferred from Fig.~\ref{fig:diskM100} that in both runs two of the sink particles create a very tight binary with a separation of about 1 AU (not resolved in the figure). The third sink in the disc is at a larger distance of about 10 AU.

\begin{figure}
 \centering
 \includegraphics[width=82mm]{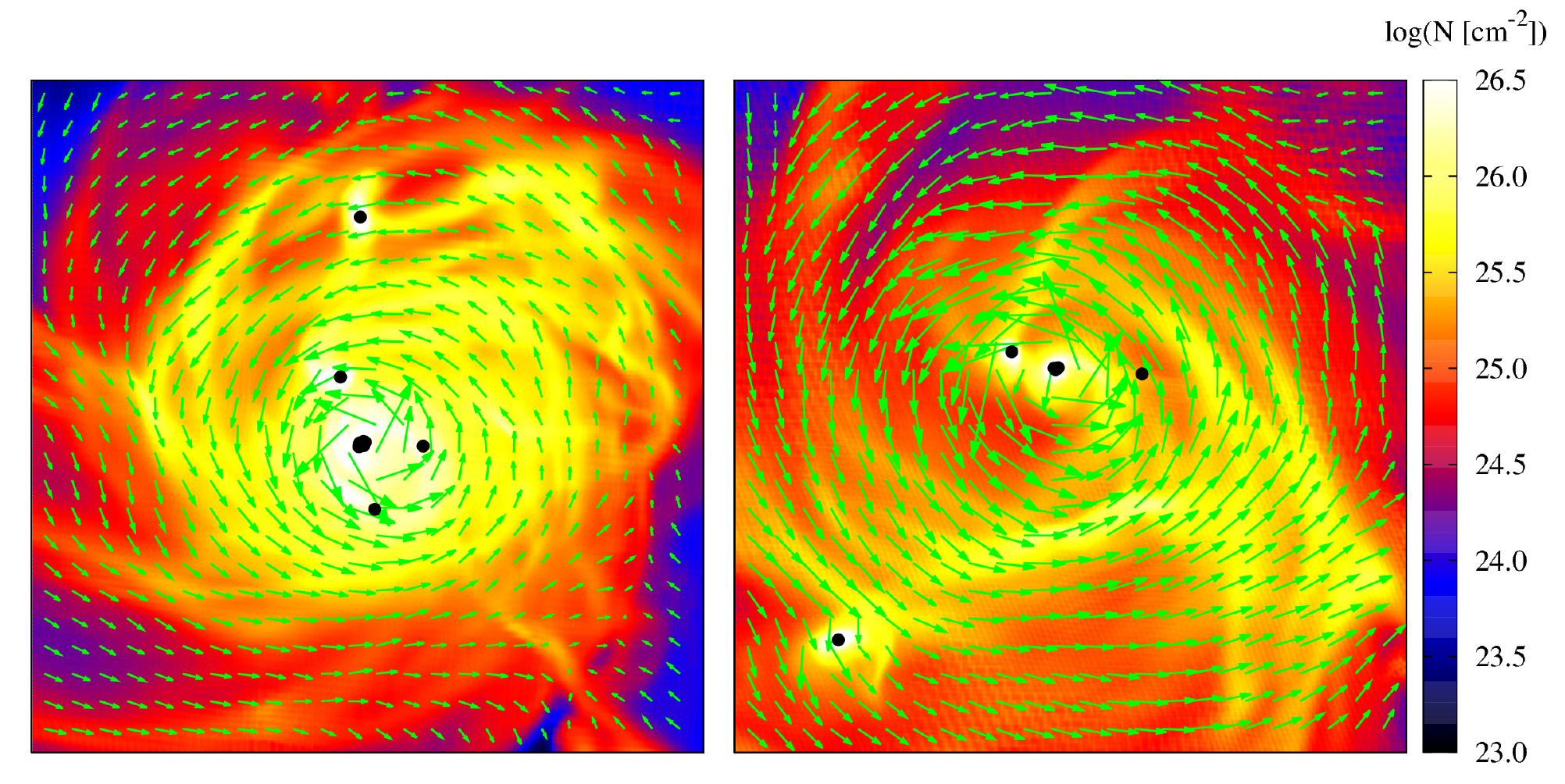}
 \includegraphics[width=82mm]{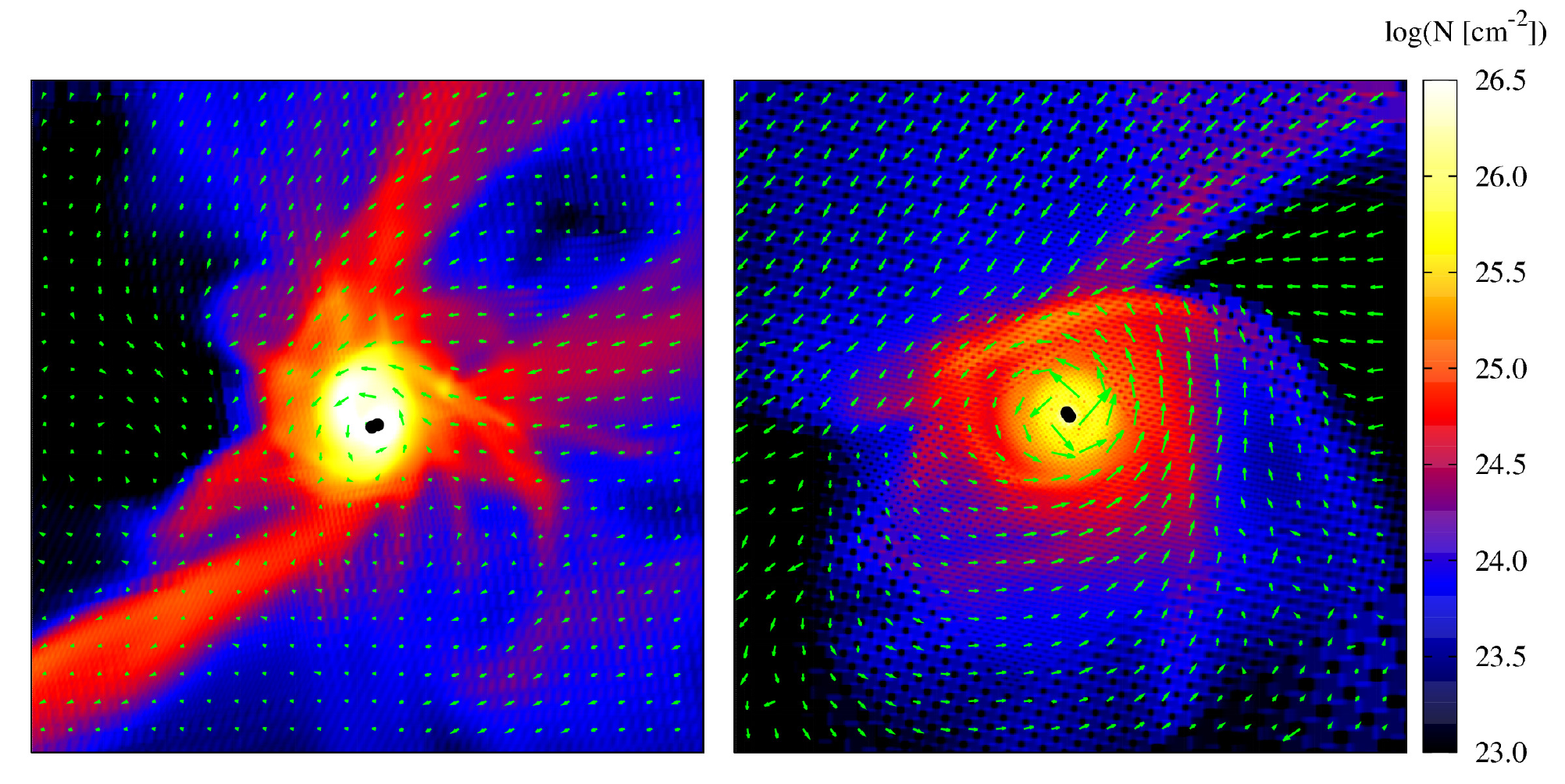}
\caption{Same as in Fig.~\ref{fig:diskM2} but for the two most massive discs (top and bottom row) in run 2.6-NoRot-M300 and times of 5 kyr and 10 kyr after the formation of the first sink particle.}
 \label{fig:diskM300}
\end{figure}

In Fig.~\ref{fig:diskM300} we consider run 2.6-NoRot-M300, which has a highly gravitationally unstable configuration and strongly supersonic turbulence with a rms-Mach number of about 5. Hence, it is not surprising that run 2.6-NoRot-M300 shows heavy fragmentation with more than 50 sink particles having been formed so far. Four discs can be found, of which we plot the two most massive ones at a time of 5 kyr and 10 kyr.
As can be seen, the first disc (top panel of Fig.~\ref{fig:diskM300}) has been subject to massive fragmentation. It contains 8 sink particles and shows several very prominent spiral arms. 5 of the 8 sinks form a very tight and strongly bound system with a size of a few AU. Despite the strongly fragmented structure the disc still retains a well-defined rotation structure. By the end of the simulation the second disc is by far smaller than the first one although also in this case a rotationally supported structure is recognisable up to a radius of about 25 AU. Furthermore, it shows significantly less fragmentation than the first disc and only contains 2 sink particles.

\begin{figure}
 \centering
 \includegraphics[width=82mm]{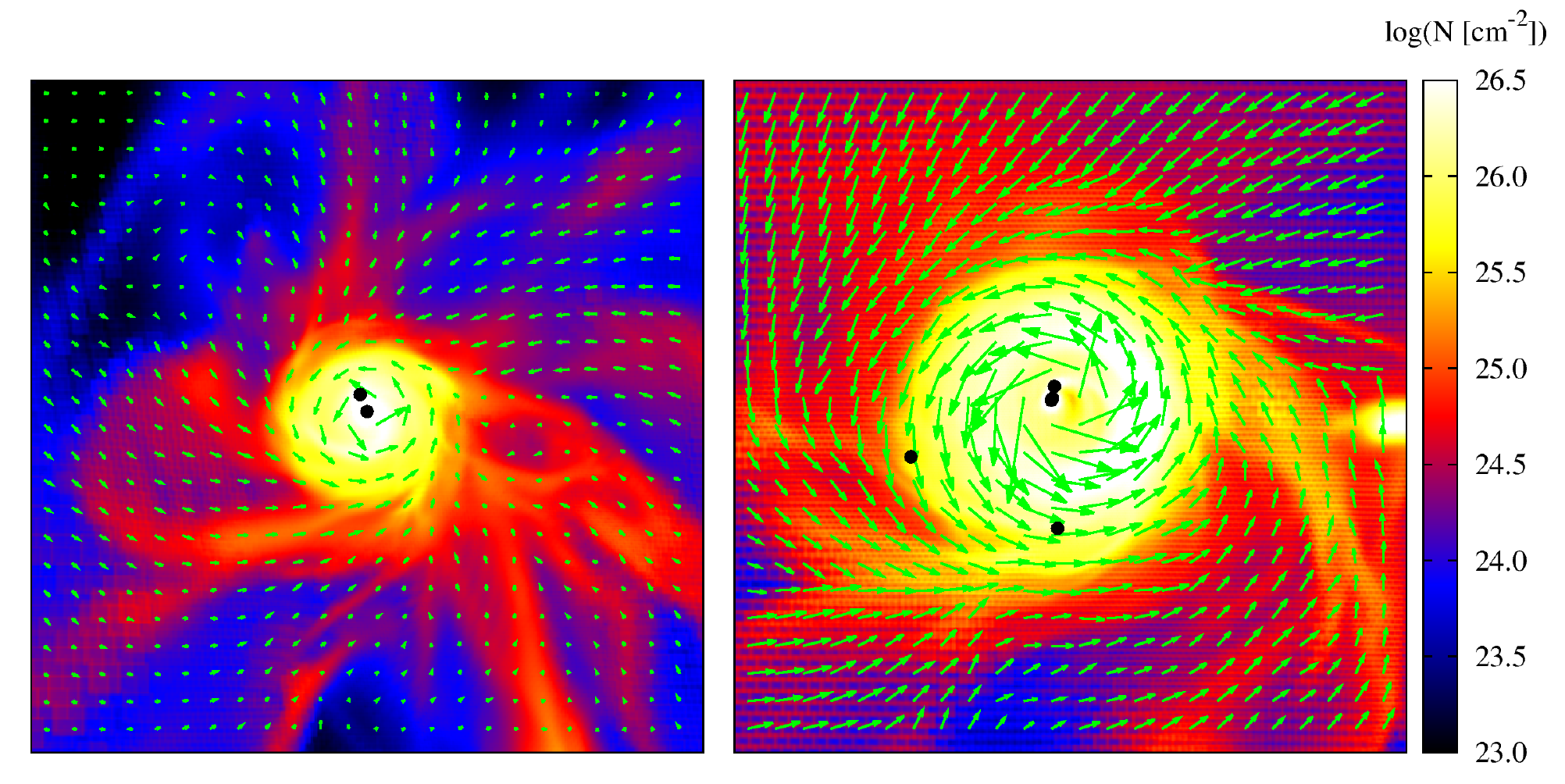}
 \includegraphics[width=82mm]{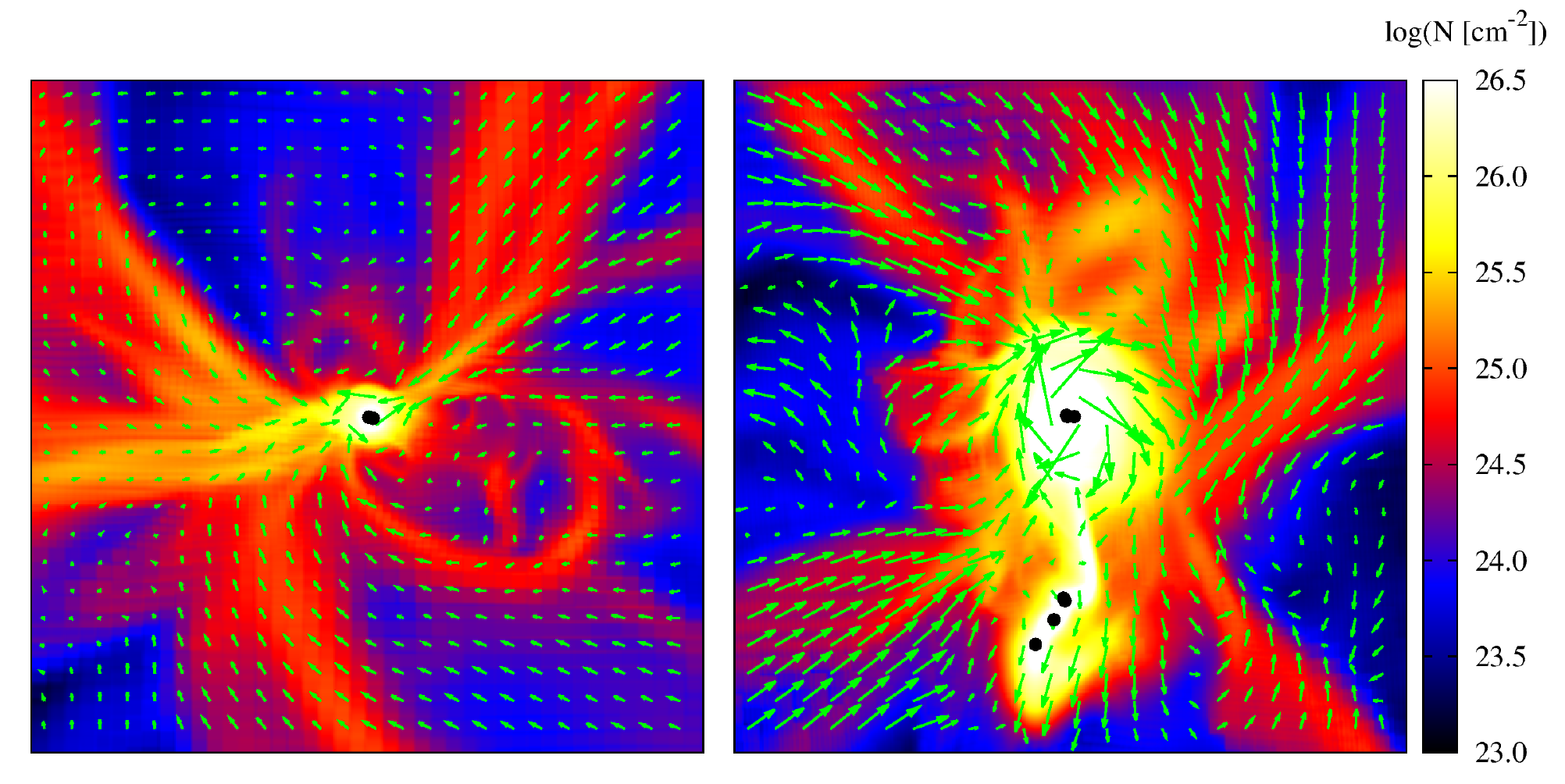}
\caption{Same as in Fig.~\ref{fig:diskM300} but for the two most massive discs (top and bottom row) in run 2.6-Rot-M1000.}
 \label{fig:diskM1000}
\end{figure}

Finally, in Fig.~\ref{fig:diskM1000} we show the two most massive discs formed in run 2.6-Rot-M1000. As in run 2.6-NoRot-M300 the 1000 M$_{\sun}$ core has fragmented heavily forming 23 sink particles by the end of the simulation. The somewhat lower degree of fragmentation (about a factor of 2) is most likely a consequence of the much more compact configuration with a $\sim$ 4 times higher central density in run 2.6-NoRot-M300. The Jeans masses given in Section~\ref{sec:techniques} (340 in run 2.6-Rot-M1000 compared to 290 in run 2.6-NoRot-M300) are only globally averaged quantities, i.e order-of-magnitude estimates, which can therefore not be taken as an absolute indication of the expected degree of fragmentation.
As can be seen, the most massive disc (top panel of Fig.~\ref{fig:diskM1000}) is significantly less fragmented than the most massive disc in run 2.6-NoRot-M300 (top panel of Fig.~\ref{fig:diskM300}), containing only 5 sinks. A ring with high column density appears at a distance of 20 -- 30 AU from the disc centre which might probably fragment in the future. The second disc (bottom panel of Fig.~\ref{fig:diskM1000}) has not yet formed around 5 kyr and only at the end of the simulation a rotating structure up to a radius of about 25 AU becomes visible.

Two important results can be inferred from the Figs.~\ref{fig:diskM2} - \ref{fig:diskM1000}. First, for all discs the typical radius, i.e. the radius where the column density reveals a sharp drop-off, is in general not larger than $\sim$ 50 -- 75 AU (except maybe the first disc in run 2.6-NoRot-M300). This is significantly smaller than the size of the rotationally supported discs we find in similar runs without turbulence~\citep{Seifried11}. In this work we find Keplerian discs with radii up to a few 100 AU although for an initial mass-flux-ratio larger by about a factor of 5 or more than that in the simulations presented here. A second important result inferred from the column density plots is the fact that all discs seem to build up a rotationally supported structure already during the earliest phase of their evolution. Indeed, we checked that in all runs rotationally supported structures become visible already around 5 kyr after the formation of the first sink particle, thus at an even earlier time than that shown in Figs.~\ref{fig:diskM2} and~\ref{fig:diskM100}.

Some global properties of the discs are listed in Table~\ref{tab:prop}. For the mass of the disc the density threshold criterion mentioned in the beginning is used. The radius of the disc is defined as the distance from the centre-of-mass where the azimuthally averaged column density drops below a value of $1 \cdot 10^{25}$ cm$^{-2}$. Although this choice is somewhat arbitrary, we found it to be reasonable from visually inspecting the column density plots. Furthermore, we only denote the accumulation of dense gas around a sink particle as a disc in case a clear rotationally supported structure larger than about 30 AU is recognisable.
\begin{table}
 \caption{Simulation properties at the end of each run: total number of sinks and discs, diameter, mass, and number of fragments in the most massive disc and the mass-to-flux ratio in its vicinity.}
 \label{tab:prop}
 \begin{tabular}{@{}lcccccc}
  \hline
  Run & N$_\rmn{sinks}$ & N$_\rmn{discs}$ & \O & $m_\rmn{disc}$ & N$_\rmn{frag}$ & $\mu$ \\
      & & & [AU] & [M$_{\sun}$] & & \\
  \hline
  2.6-NoRot-M2 & 1 & 1 & 67 & 0.0585 & 1 & 6.2 \\
  2.6-Rot-M2 & 1 & 1 & 89 & 0.0607 & 1 & 6.6 \\
  2.6-NoRot-M100 & 5 & 1 & 124 & 0.175 & 3 & 12.9 \\
  2.6-Rot-M100 & 5 & 1 & 124 & 0.452 & 3 & 14.4 \\
  2.6-Rot-M100-B & 36 & 4 & 71 & 0.062 & 4 & 11.5 \\
  2.6-Rot-M100-C & 5 & 1 & 111 & 0.263 & 3 & 9.3 \\
  2.6-Rot-M100-p2 & 5 & 1 & 193 & 0.63 & 4 & 14.6 \\
  2.6-NoRot-M300 & 57 & 4 & 132 & 0.048 & 8 & 9.2 \\
  2.6-Rot-M1000 & 23 & 4 & 109 & 0.40 & 5 & 9.7 \\
  \hline
 \end{tabular}
\end{table}

\subsubsection{Velocity structure}

Next, we consider the velocity structure in the discs in more detail. For this purpose we calculate the rotation velocity $v_\phi$ and the radial velocity $v_\rmn{rad}$ for each cell in the disc. We note that the calculation is done in the rest-frame of the discs, i.e. the $z$-axis is defined by the disc angular momentum vector and the velocity is corrected by the velocity of the disc's centre-of-mass before calculating $v_\phi$ and $v_\rmn{rad}$. In contrast to the column density plots presented before we restrict the consideration to the final snapshot at the end of each simulation. However, we emphasise that when considering the velocity structure at earlier times, the results do not change qualitatively as could already be inferred from the velocity vectors shown in the column density plots. Furthermore, we note that for radii larger than $\sim$ 50 AU (depending on the actual simulation), where no cells with gas densities above $5 \cdot 10^{-13}$ g cm$^{-3}$ are found, we adopt a simple geometrical criterion considering all the gas within a height of 20 AU above/below the midplane defined by the disc. Hence, we are able to analyse the velocity structure also outside the disc defined by the density threshold. In order to get an impression of the scatter of $v_\phi$ and $v_\rmn{rad}$, the values are not averaged azimuthally. The Keplerian velocity $(GM/r)^{1/2}$ is calculated by taking into account the mass of gas and all sink particles within a sphere of radius $r$ around the centre of the disc.

\begin{figure}
 \centering
 \includegraphics[width=41mm]{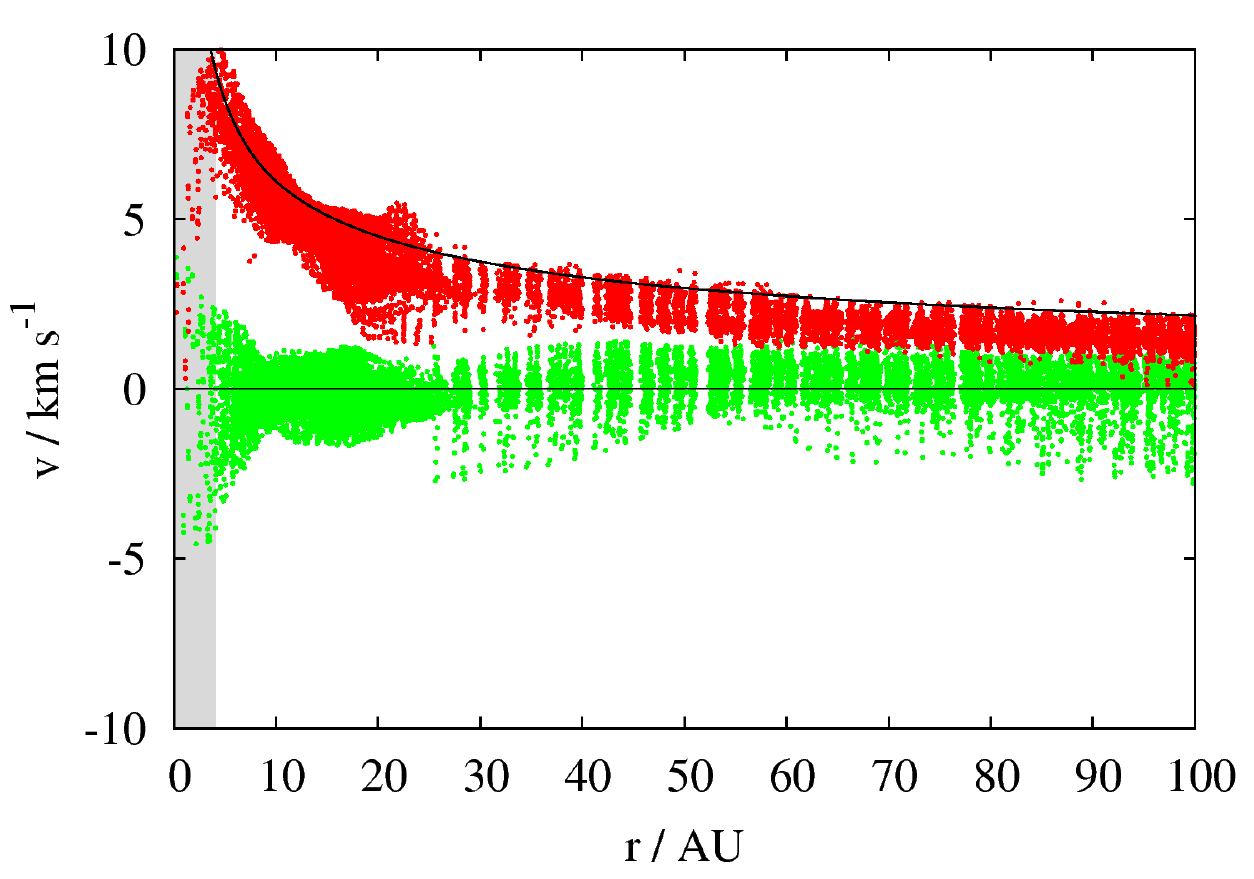}
 \includegraphics[width=41mm]{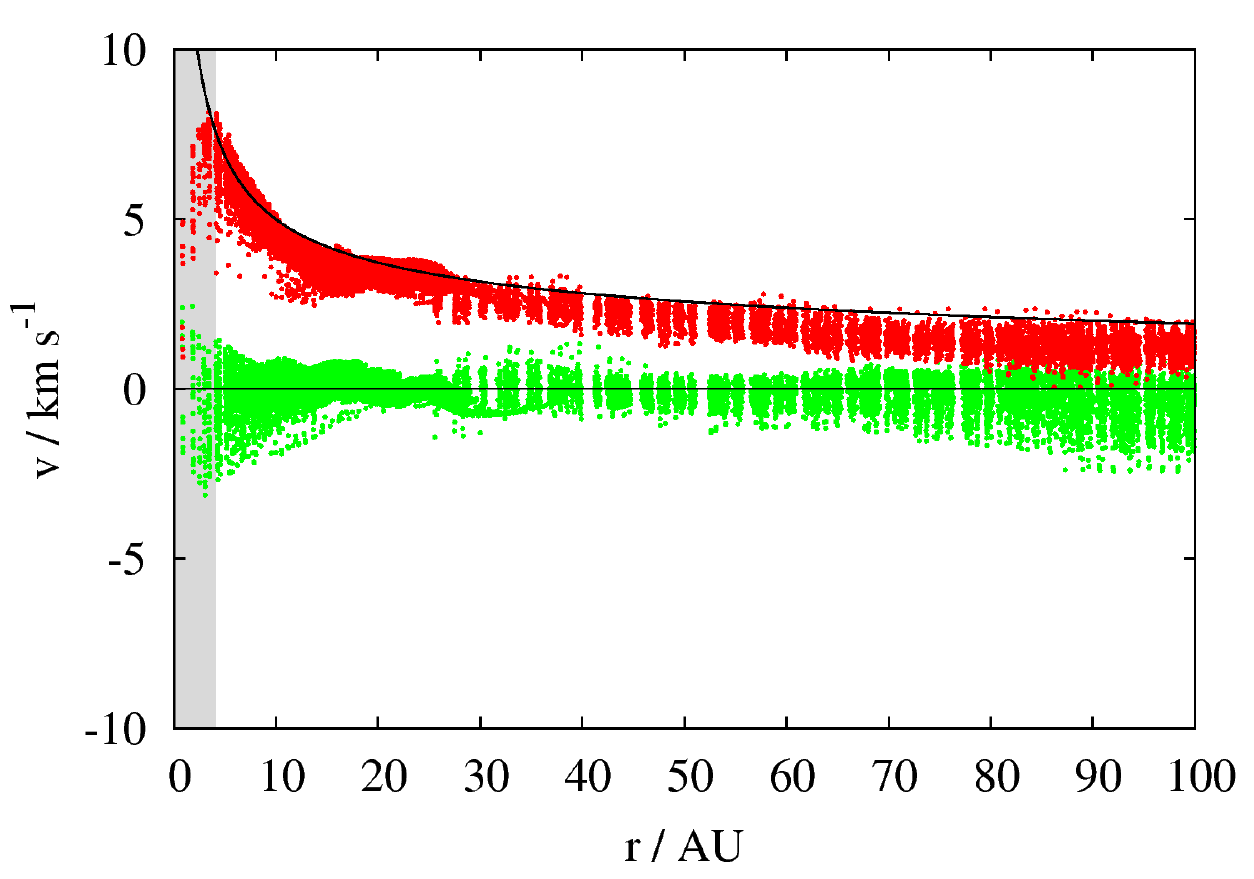}
 \includegraphics[width=41mm]{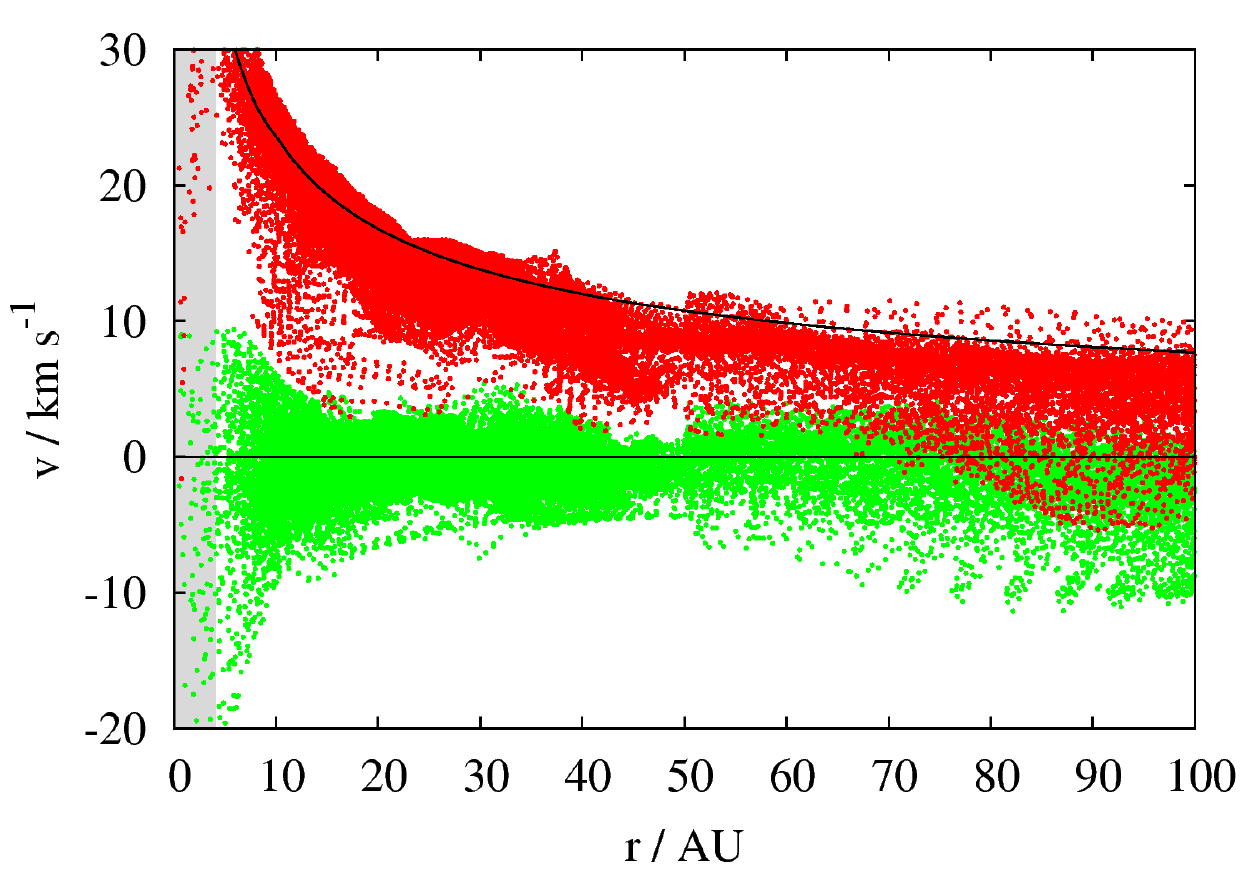}
 \includegraphics[width=41mm]{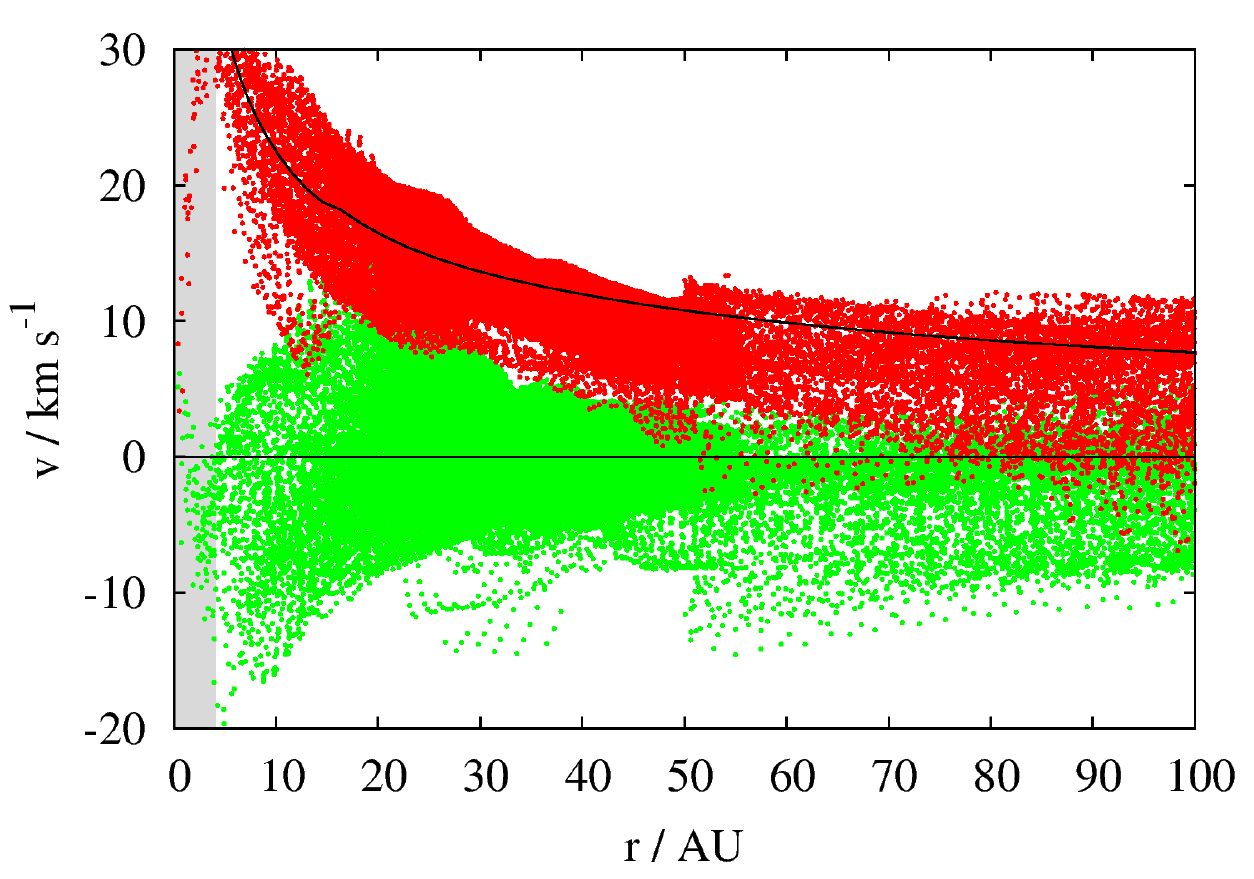}
 \caption{Radial dependence of the rotation (red) and radial velocity (green) for the four discs of the runs 2.6-NoRot-M2, 2.6-Rot-M2, 2.6-NoRot-M100 and 2.6-Rot-M100 (from top left to bottom right) at the end of each simulation. The black solid line shows the Keplerian velocity $v_\rmn{kep}$. The regions below 4 AU are affected by resolution effects, therefore they are shaded grey to guide the reader's eye. Note the different scaling of the y-axis.}
 \label{fig:velM2-100}
\end{figure}

\begin{figure}
 \centering
 \includegraphics[width=41mm]{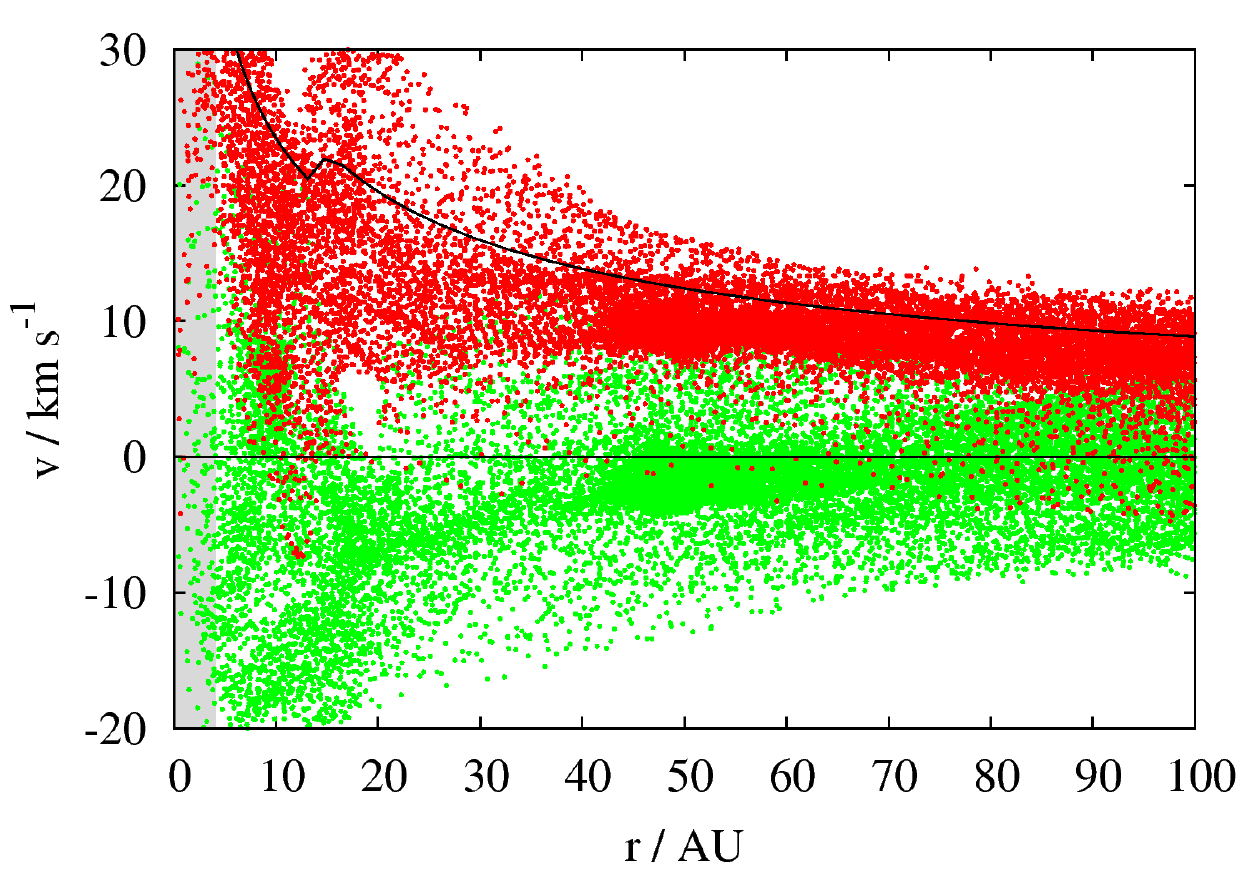}
 \includegraphics[width=41mm]{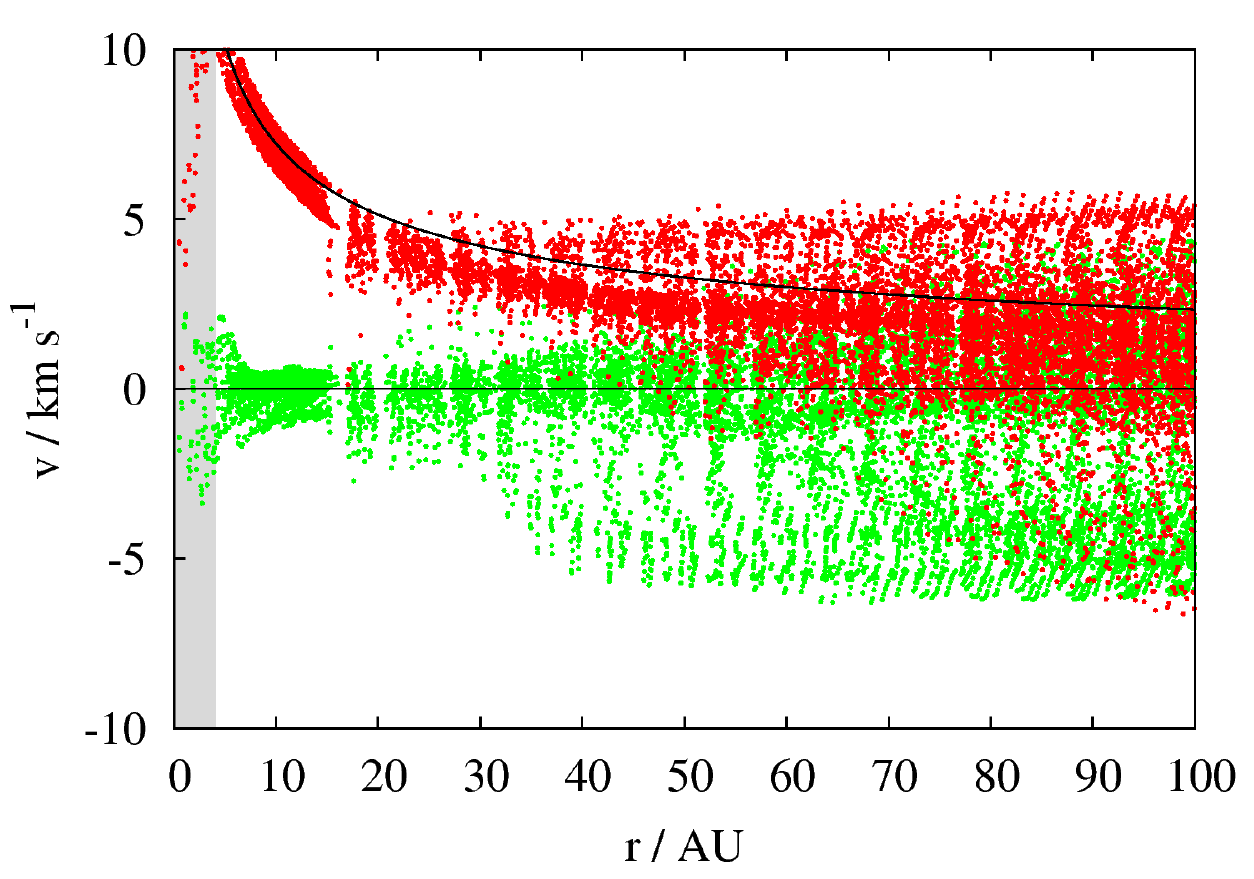}
 \includegraphics[width=41mm]{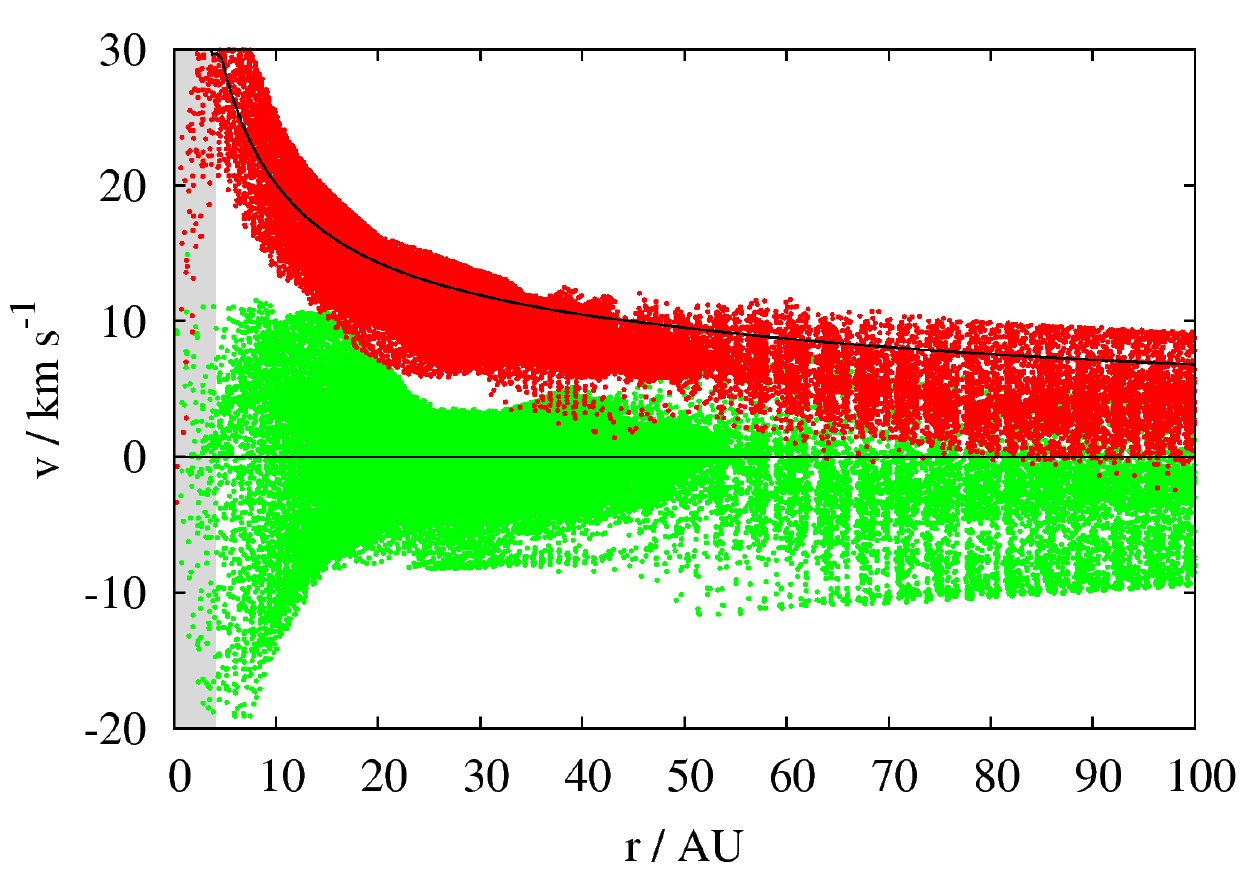}
 \includegraphics[width=41mm]{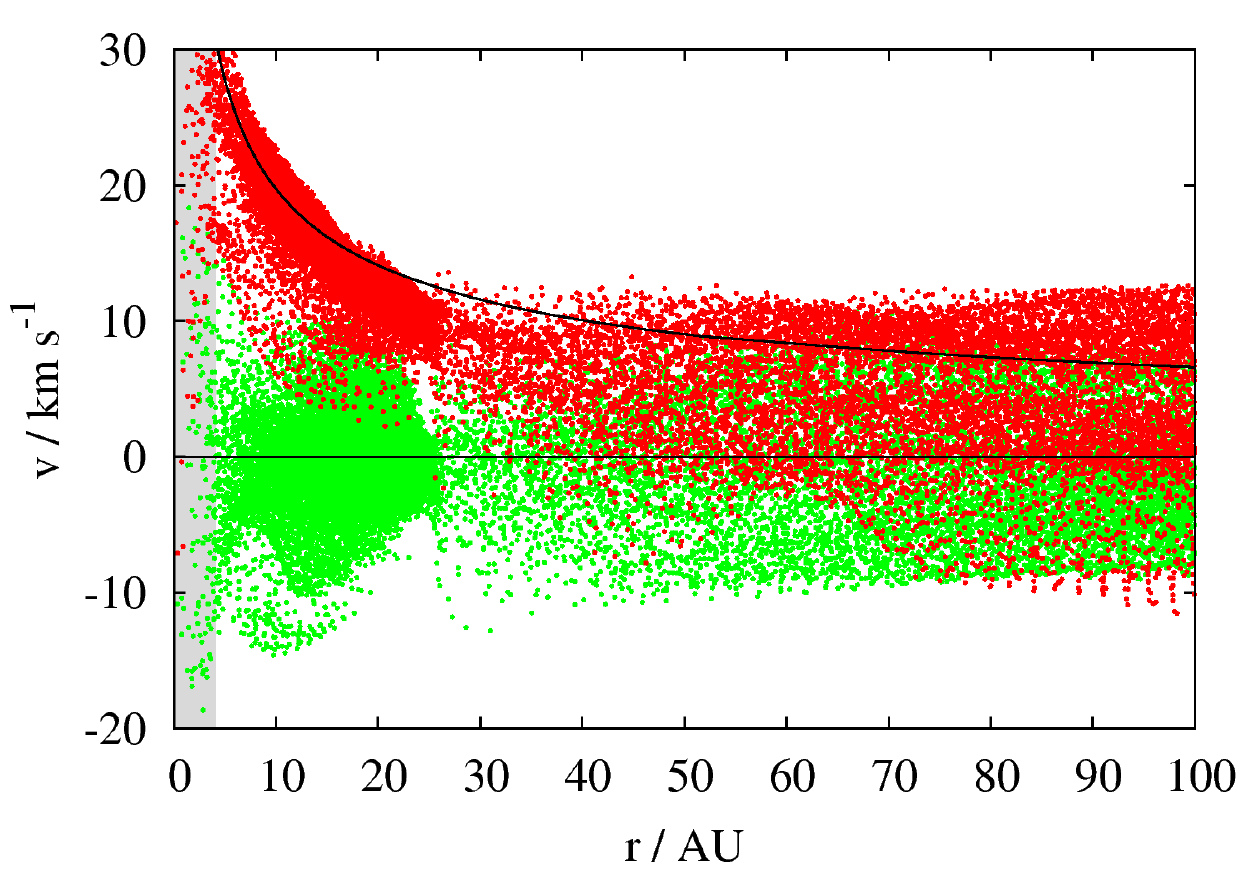}
 \caption{Same as in Fig.~\ref{fig:velM2-100} but for the two most massive discs of the runs 2.6-NoRot-M300 (top) and 2.6-Rot-M1000 (bottom).}
 \label{fig:velM300-1000}
\end{figure}

In Fig.~\ref{fig:velM2-100} we plot the radial dependence of $v_\phi$ and $v_\rmn{rad}$ for the four discs in the runs 2.6-NoRot-M2, 2.6-Rot-M2, 2.6-NoRot-M100 and 2.6-Rot-M100 and in Fig.~\ref{fig:velM300-1000} for the two most massive discs in the runs 2.6-NoRot-M300 and 2.6-Rot-M1000.
It can be seen that within the typical disc radius inferred from the column density plots ($\la$ 50 AU) the discs in general rotate with the Keplerian velocity (black lines). The radial velocity on the other hand scatters around zero as expected for a Keplerian disc and is in general significantly smaller than the rotation velocity. Interestingly, for some of the runs the Keplerian velocity structure extends up to $\sim$ 100 AU and thus even beyond the disc radius inferred from the column density plots. This result is in strong contrast to a series of star formation simulations in which turbulence is neglected~\citep[e.g.][]{Allen03,Price07,Mellon08,Hennebelle08,Duffin09,Seifried11}. In general, in these simulations it was found that for mass-to-flux ratios below 5 - 10, i.e. for moderately strong magnetic fields, no rotationally supported discs are formed. Again, we stress that these results are due to the particular choice of the initial conditions namely the omission of turbulence. On the other hand, if turbulence is taken into account, recently the build-up of early-type Keplerian discs was reported even for strong magnetic fields~\citep{Seifried12,Santos12b,Santos12}.

As demonstrated in this section, the build-up of Keplerian discs does not depend on the mass of the prestellar core -- and hence also not on the strength of the turbulence, which ranges from subsonic motions for the low-mass cores to highly supersonic motions for the most massive cores. Furthermore, for the 2.6 M$_{\sun}$ and 100 M$_{\sun}$ cores it can be seen that the lack of an overall core rotation does not significantly affect the build-up of Keplerian discs since turbulence always produces local shear flows which carry a sufficient amount of angular momentum to build up the discs. For both cases, the rotating and the non-rotating, Keplerian discs build up in the low-mass and high-mass case. However, a close inspection Fig.~\ref{fig:diskM2} and~\ref{fig:diskM100} shows that in the case of an overall core rotation the discs seem to be slightly larger than for the non-rotation case. This statement, however, has to be taken with caution since we obviously require a larger number of discs in order to obtain statistically significant results. Nevertheless, our result of slightly larger discs for uniformly rotating cores agrees with what one would naively expect.

To summarise, the mechanism of turbulence-induced disc formation as proposed recently in~\citet{Seifried12} works for a wide range of core masses, turbulence strengths, and even for cores without uniform initial rotation. We again emphasise that the Keplerian disc structure seen in Figs.~\ref{fig:velM2-100} and~\ref{fig:velM300-1000} at 15 kyr is present at much earlier times already, i.e. as early as 5 kyr. Assuming rotation velocities of a few km s$^{-1}$ at a radius of 100 AU, this time corresponds to about one to a few rotation periods of the disc -- and correspondingly more at smaller radii-- indicating that Keplerian discs build up in the very beginning. In the following we will explore this mechanism and the differences to the non-turbulent case in more detail.

\subsection{Flux loss}
\label{sec:massflux}

The extensive work on disc formation under the influence of magnetic fields has revealed a critical value for the mass-to-flux ratio of 5 -- 10 below which the formation of Keplerian discs is suppressed in collapsing cloud cores without turbulent motions~\citep[e.g.][]{Allen03,Price07,Mellon08,Hennebelle08,Duffin09,Seifried11}. In the following we test to what extent the formation of Keplerian discs in our simulations can be explained by the loss of magnetic flux in the discs' vicinity. For this purpose we calculate the mass-to-flux ratio $\mu$ in spheres of radius $r$ = 500 AU around the centre-of-mass of the discs formed in the four runs 2.6-NoRot-M2, 2.6-Rot-M2, 2.6-NoRot-M100 and 2.6-Rot-M100. We note that we have chosen a radius of 500 AU for a particular reason: By calculating the gas torque and the magnetic torque in~\citet{Seifried11} we have shown that a large fraction of the angular momentum is removed already \textit{before} the gas falls onto the disc. The analysis showed that up to a distance of 500 -- 1000 AU from the centre of the disc the (negative) magnetic torque balances the (positive) inwards angular momentum transport by the gas~\citep[see bottom panel of Fig. 11 in][]{Seifried11}. Hence, this result clearly illustrates the importance to check what happens on scales larger than 100 AU.

\begin{figure}
 \centering
 \includegraphics[width=60mm]{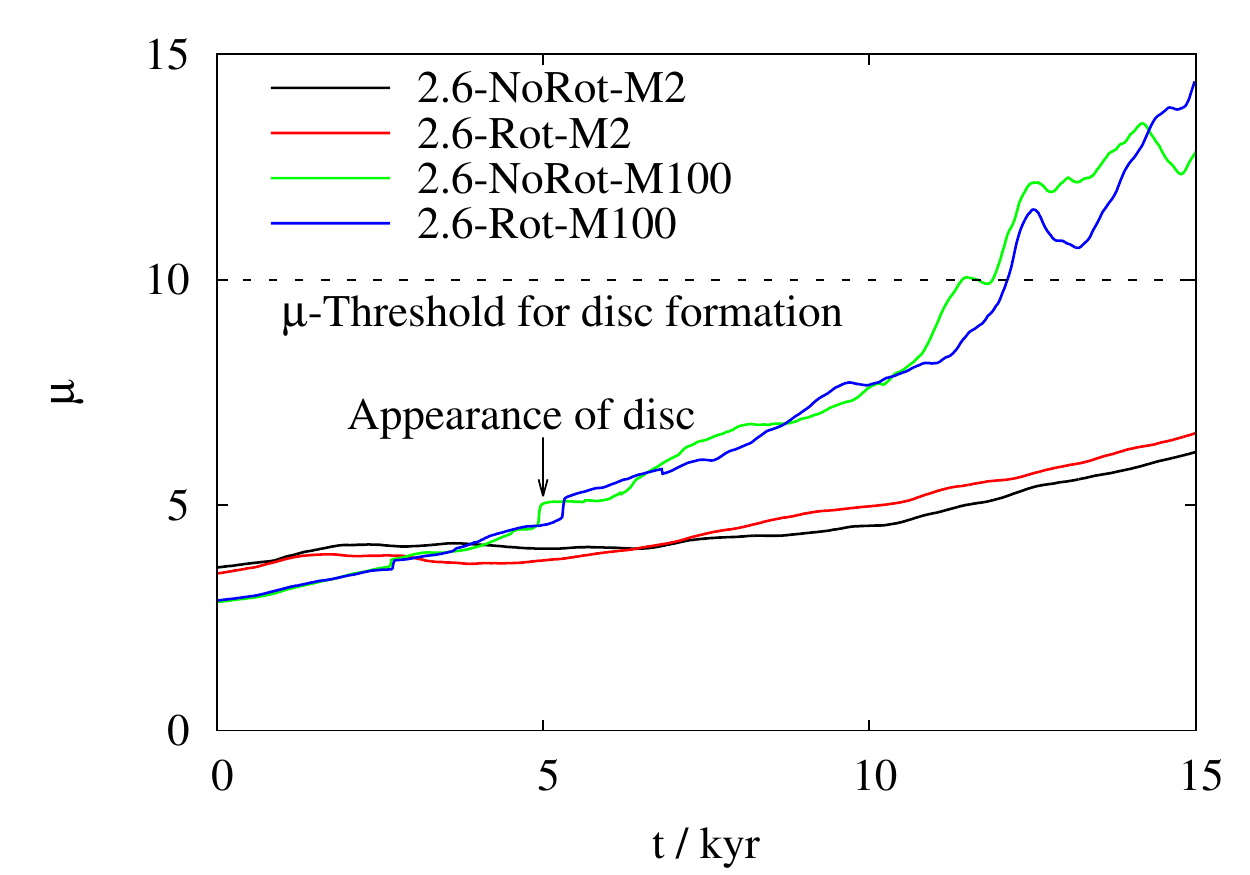}
 \includegraphics[width=60mm]{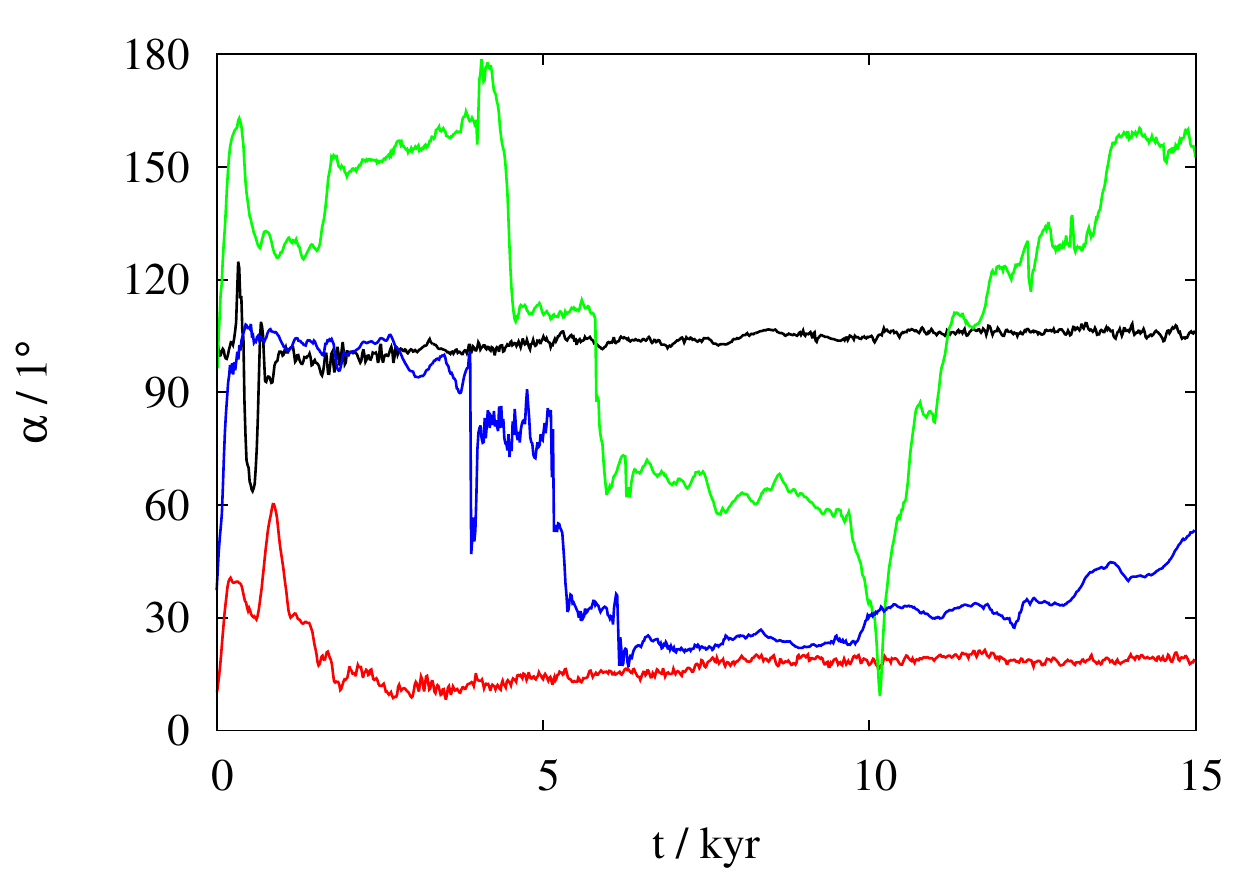}
 \caption{Top: Mass-to-flux ratio for spheres with a radius of 500 AU around the discs of the runs 2.6-NoRot-M2, 2.6-Rot-M2, 2.6-NoRot-M100 and 2.6-Rot-M100. The straight, black, dashed line indicates the threshold for $\mu$ below which Keplerian disc formation is suppressed. The arrow indicates the time when a disc with significant rotational support is observed for the first time. Bottom: Inclination of the disc angular momentum to the mean magnetic field for the same spheres as in the top panel.}
 \label{fig:muM2-100}
\end{figure}

In the top panel of Fig.~\ref{fig:muM2-100} we plot $\mu$ in the $r$ = 500 AU-sphere.
It can be seen that $\mu$ increases over time reaching values $> 10$ for the 100 M$_{\sun}$ runs. This indicates that there is a loss of magnetic flux in the vicinity of the discs over time by a factor of a few. However, we emphasise that a part of the increase of $\mu$ is most likely due to accretion along the magnetic field lines -- a fact which cannot be covered by a simple geometrical calculation of the mass-to-flux ratio and which is often ignored in literature. Hence, it is likely that the influence of flux loss e.g. by magnetic reconnection~\citep{Santos12b,Santos12} on the increase of $\mu$ and thus the formation of Keplerian discs is somewhat overestimated.

Despite the steady increase of $\mu$ the observed flux loss is still too small to explain the build-up of Keplerian discs on its own. For the 100 M$_{\sun}$ runs $\mu$ exceeds 10 only after 10 kyr, i.e. well \textit{after} the Keplerian disc has built up, and for the 2.6 M$_{\sun}$ runs it stays below 10 for the entire 15 kyr. Since non-turbulent simulations, however, suggest that a value of $\mu > 10$ is necessary for Keplerian discs to form, clearly the loss of magnetic flux alone cannot explain the observed build-up of Keplerian discs at t $\la$ 5 kyr. We note that for the remaining runs not considered here the time evolution of $\mu$ is qualitatively and quantitatively very similar to the runs 2.6-NoRot-M100 and 2.6-Rot-M100, i.e. $\mu$ reaches values $\ga$ 10 at the end of the simulation (see Table~\ref{tab:prop}).

\subsection{Magnetic field structure}

Recently, \citet{Hennebelle09},~\citet{Ciardi10} and~\citet{Joos12} suggested that an inclination of the magnetic field with respect to the rotation axis of the core decreases the magnetic braking efficiency. In order to test this statement, we inspected the magnetic field structure in the vicinity of our discs. For this purpose, in the bottom panel of Fig.~\ref{fig:muM2-100} we plot the inclination $\alpha$ of the disc angular momentum to the average magnetic field in the sphere of $r$ = 500 AU. The inclination $\alpha$ is found to be relatively large except for run 2.6-Rot-M2. We note that also for the remaining runs there is in general a significant inclination between the disc angular momentum and the mean magnetic field. Interestingly, $\alpha$ seems to be somewhat smaller for the runs with overall rotation. This can be easily understood given the fact that for these runs the disc angular momentum at later times is much closer to the $z$-axis (see left panel of Fig.~\ref{fig:theta}) and thus also closer to the overall direction of the magnetic field in the core. 

\begin{figure}
 \centering
 \includegraphics[width=60mm]{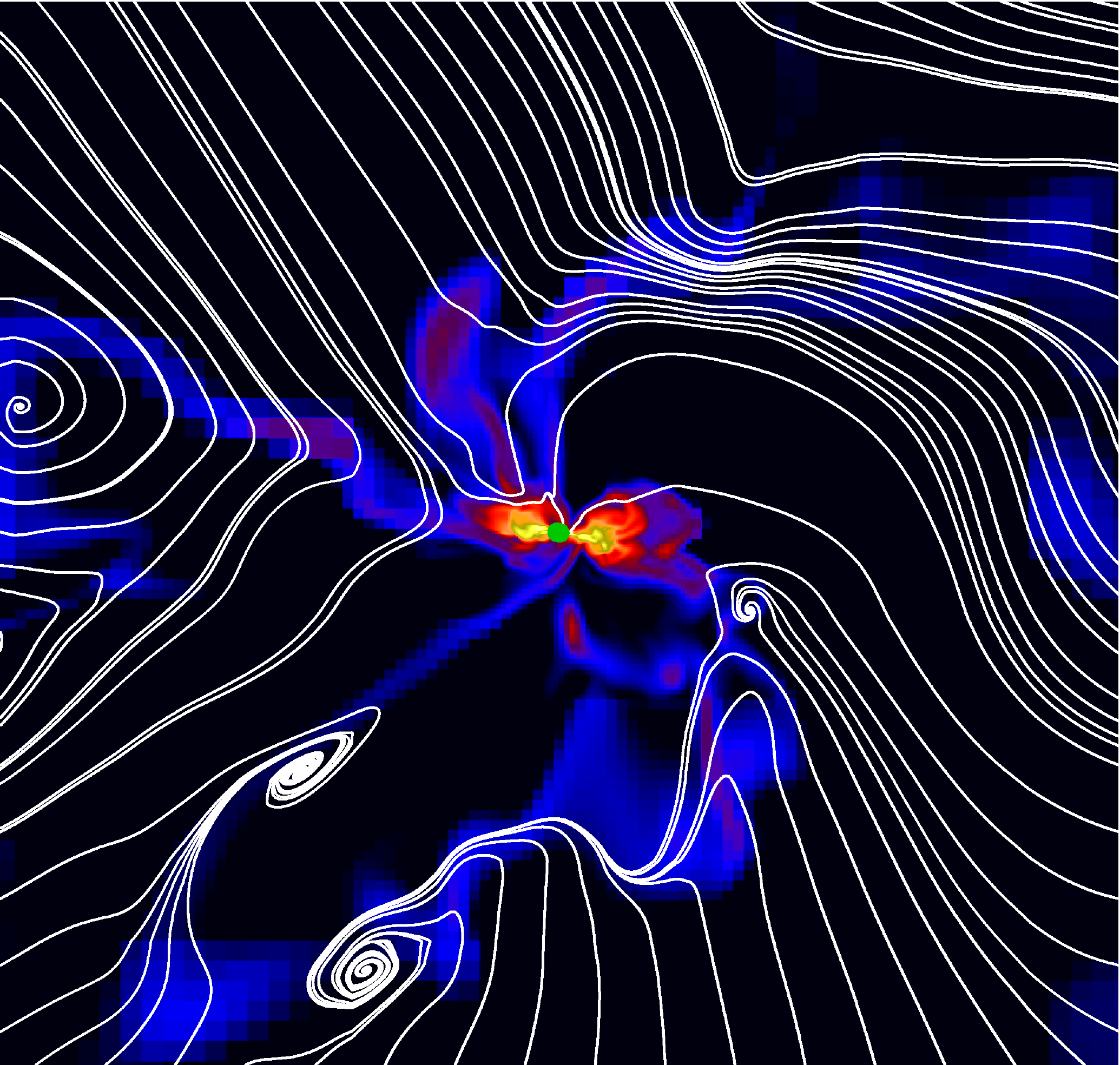}
 \caption{Edge-on view of the disc in run 2.6-Rot-M100. The column density is overplotted with the magnetic field lines (white) and the sink particles (green dots). The figure is 1000 AU in size thus showing the same region analysed in Fig.~\ref{fig:muM2-100}.}
 \label{fig:edgeon}
\end{figure}

At first sight, the relatively large inclination between the angular momentum of the disc and the mean magnetic field support the findings of \citet{Hennebelle09},~\citet{Ciardi10} and~\citet{Joos12} showing that inclined magnetic field lines aid the formation of Keplerian discs even for high magnetic field strengths. To investigate those arguments in more detail, we inspect the structure of the magnetic field lines in the vicinity of the disc in run 2.6-Rot-M100 in Fig.~\ref{fig:edgeon}.
Although a preferred direction of the magnetic field is recognisable, there is a significant distortion of the magnetic field due to the turbulence on scales of a few 100 AU. For such a distorted magnetic field, however, the definition of a mean field direction, as used by \citet{Hennebelle09},~\citet{Ciardi10} and~\citet{Joos12}, and its inclination to the disc orientation can be considered questionable. Hence, we strongly suggest that the large inclination between the disc orientation and the mean magnetic field measured in the bottom panel of Fig.~\ref{fig:muM2-100} is most likely not the only reason for the build-up of Keplerian discs. 

The distortion of the magnetic field in the surroundings of the discs itself can explain the reduced magnetic braking efficiency and the subsequent build-up of Keplerian discs since the disordered field structure prevents an effective coupling of the inner parts close to the disc to the outer parts. A magnetic field line structure, however, which fans out from a typical disc radius $R_0$ to a radius $R$ somewhere outside, would significantly increases the magnetic braking efficiency by a factor of $(R/R_0)^4$~\citep{Mouschovias85}. For a non-turbulent simulation usually such a configuration is present, see e.g. Fig. 13 in~\citet{Hennebelle08} or Fig. 9 in~\citet{Seifried11}. In our simulations, however, only a strongly disordered field structure is present due to the turbulent motions. This is what we consider to be the main reason for the reduced magnetic braking efficiency. We emphasise, however, that more generally one could regard the reduced magnetic braking efficiency due to misaligned magnetic fields~\citep{Hennebelle09,Ciardi10,Joos12} as being closely related to our case when interpreting the misalignment as some kind of disorder.

Moreover, the magnetic braking mechanism as described by~\citet{Mouschovias80} relies on the build-up of a toroidal magnetic field due to coherent rotational motions. In the vicinity of the discs presented here such coherent rotation structures are barely present. Instead, mass accretion occurs along distinct channels (shear flows) created locally by the turbulent motions (see Figs.~\ref{fig:diskM2} -- \ref{fig:diskM1000}). Since we have shown in~\citet{Seifried11} that a significant fraction of the angular momentum is removed by magnetic braking already before the gas falls onto the disc, the lack of a coherent rotation structure in the surroundings of the discs further reduces the magnetic braking efficiency. Simultaneously, the shear flows can provide enough angular momentum to build up the discs as already pointed out in~\citet{Seifried12}.

\subsection{Time evolution}
\label{sec:time}

For the analysis of the time evolution of the discs we restrict ourselves to the runs 2.6-NoRot-M2, 2.6-Rot-M2, 2.6-NoRot-M100 and 2.6-Rot-M100. There are some interesting differences between low-mass and high-mass cores as well as rotating and non-rotating cores, which will be demonstrated in the following. We remind the reader that we define the discs by all the gas with a density larger than $5 \cdot 10^{-13}$ g cm$^{-3}$ around the corresponding sink particle.

\begin{figure*}
 \centering
 \includegraphics[width=58mm]{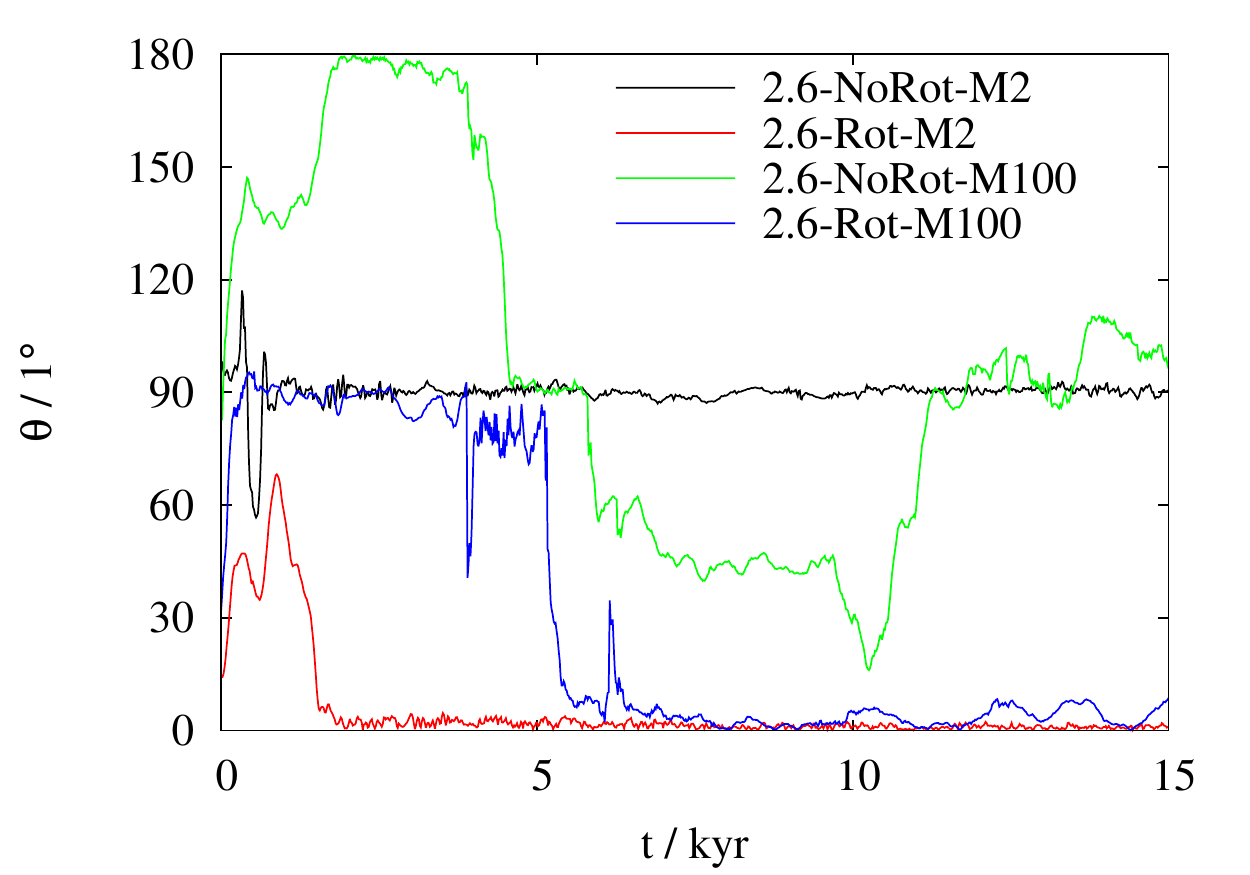}
 \includegraphics[width=58mm]{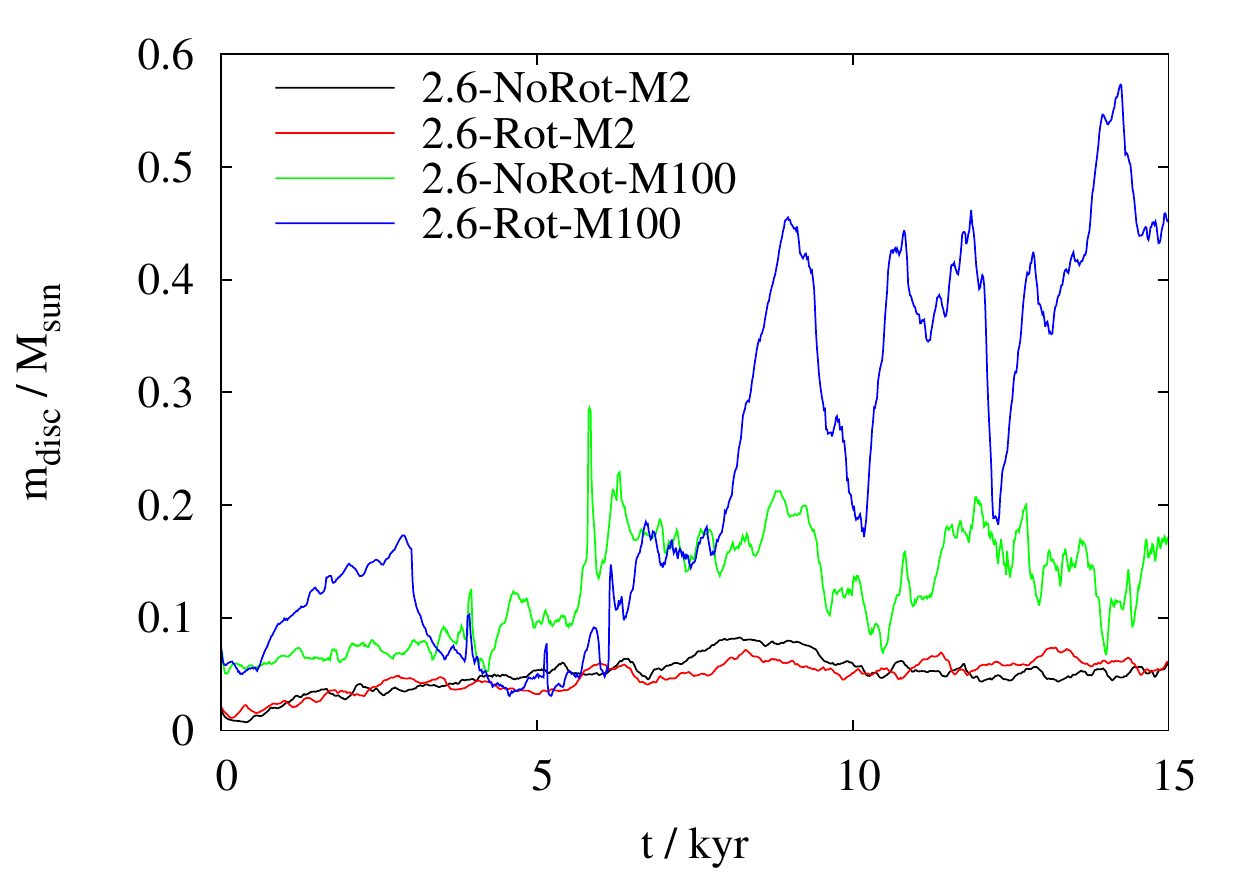}
 \includegraphics[width=58mm]{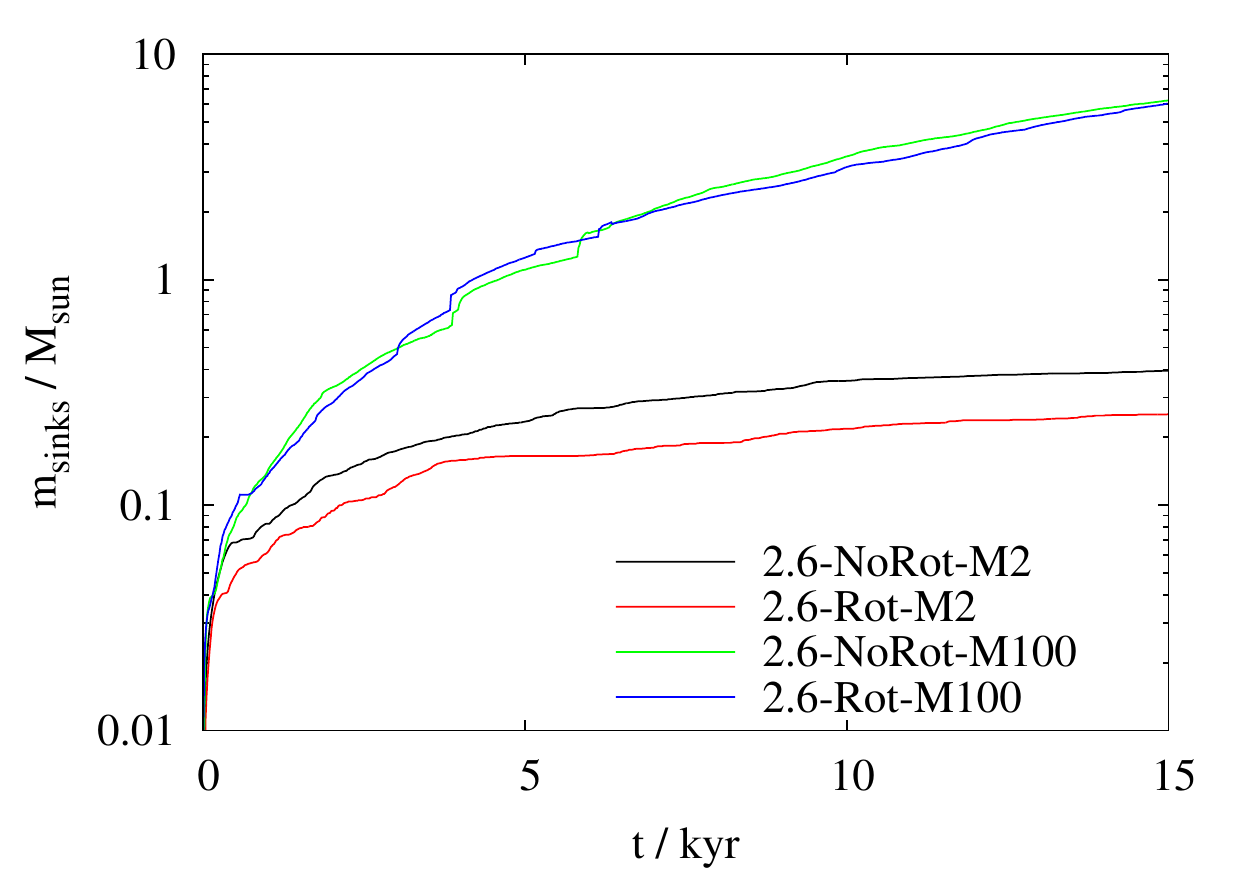}
 \caption{Left: Time evolution of the disc orientation as measured by the angle between the $z$-axis and the disc angular momentum. For the runs with initial uniform rotation the disc's angular momentum clearly approaches the overall angular momentum over time. Middle: Time evolution of the disc masses. Right: Time evolution of the total mass of all sinks contained in the disc.}
 \label{fig:theta}
\end{figure*}

First, we consider the time evolution of the orientation of the discs in the left panel of Fig.~\ref{fig:theta}. For this purpose we calculate the angular momentum of the discs and show its deviation from the $z$-axis, i.e. the polar angle $\theta$. We have chosen the $z$-axis as it corresponds to the direction of the initial magnetic field.
As can be seen there is a significant difference between the runs with an overall core rotation compared to that without any rotation as well as between the low- and high-mass runs. In the beginning the four discs have an angular momentum well off from the $z$-axis. However, for the runs 2.6-Rot-M2 and 2.6-Rot-M100 the orientation of the disc angular momentum approaches the $z$-axis ($\theta \simeq 0$) after some time, for run 2.6-Rot-M2 significantly earlier than for run 2.6-Rot-M100. Moreover, one can see that this change in the direction of the angular momentum happens rather quickly -- within about 1000 yr. After the disc orientation has approached the $z$-axis it stay more or less constant, which is particularly interesting for run 2.6-Rot-M100 since here strongly supersonic turbulent motions are present. Such an alignment of the rotation and magnetic field direction was first observed by~\citet{Machida06}.

Interestingly, for run 2.6-NoRot-M2, where no overall rotation but only turbulent motions are present, the disc orientation remains remarkable constant and also the azimuthal angle $\phi$ does not change significantly over time. We note that we decided not to show the azimuthal angle in Fig.~\ref{fig:theta} since for the cases where $\theta$ is close to zero the variation in $\phi$ can be very large -- although the disc orientation itself does hardly change in this case -- which would make the figure basically unreadable. In contrast to run 2.6-NoRot-M2, run 2.6-NoRot-M100 shows large variations in the disc orientation, which we attribute to the very turbulent environment they form in. Again the changes in the disc's orientation occur rapidly within about 1000 yr or even less. This points to very distinct accretion events of blobs of gas onto the disc, which results in an almost instantaneous change of the angular momentum.

In run 2.6-Rot-M100 the mass of the disc is a factor of a few larger than that of the disc in run 2.6-NoRot-M100. This could be attributed to the closer orientation of the disc in run 2.6-Rot-M100 to the $z$-axis thus profiting from the large reservoir of angular momentum due to the overall rotation. For the runs with 2.6 M$_{\sun}$, however, both discs have rather similar masses of about 0.05 M$_{\sun}$. Furthermore, these discs do not grow over time but keep their mass whereas the discs in the high-mass core runs show a clear increase of mass over time despite distinct events of significant mass loss. Moreover, a close comparison between the disc orientation and mass for run 2.6-NoRot-M100 shows that the rapid change of the disc's orientation after 4 kyr and 6 kyr is correlated to a strong increase of the disc mass. This confirms our assumption of distinct accretion events onto the disc being responsible for the rapid changes. We note that at the end of each simulation the disc masses are well below the total sink particle masses in the discs, which range from 0.25 M$_{\sun}$ in run 2.6-Rot-M2 to 6.3 M$_{\sun}$ in run 2.6-NoRot-M100 (right panel of Fig.~\ref{fig:theta}). We find that phases of significant decrease in disc mass (see middle panel of Fig.~\ref{fig:theta}) are usually accompanied with increased sink accretion rates. This can be explained by the preceding accretion of gas onto the disc with a small or even negative angular momentum (with respect to the disc angular momentum). Hence, the specific angular momentum of the disc is lowered and so is the rotational support, which in turn results in a subsequent increase of the sink accretion rate until a rotationally supported equilibrium in the disc is established again.

We emphasise that the discs in the runs not discussed above show qualitatively very similar properties with typical masses of a few 0.1 M$_{\sun}$ and partly strong fluctuations (see Table~\ref{tab:prop}) and a disc angular momentum well off from the $z$-axis (and thus from the overall magnetic field). However, as for the runs 2.6-Rot-M2 and 2.6-Rot-M100 in all other runs with an overall core rotation the orientation of the disc approaches the $z$-axis over time as seen already before. The only exemption is run 2.6-Rot-M100-C, where the disc stays misaligned over the entire 15 kyr.

To summarise, we find that there are some differences in the time evolution of the discs between the runs with and without core rotation as well as between high- and low-mass cores. Nevertheless, as demonstrated in Section~\ref{sec:discs}, the velocity structure of the individual discs is barely affected.

\subsection{Fragmentation behaviour}
\label{sec:fragmentation}

Recently, it was suggested that a large fraction of low-mass protostar and brown dwarfs form via gravitational fragmentation of protostellar discs~\citep[see e.g.][and references therein]{Stamatellos11}. In contrast,~\citet{Bate09} and~\citet{Offner09} find that the degree of disc fragmentation is reduced when including radiative feedback. In these simulations low-mass stars mainly form via the collapse of distinct gravitational cores or condensations~\citep[see also][]{Padoan04,Hennebelle08b}. At first sight the multiple sink particle systems in the discs in our high-mass core runs support the classical disc fragmentation picture~\citep{Toomre64}. However, a second mechanism to form such a system is the so-called disc-assisted capture~\citep[see e.g. Section 5.4 in][and references therein]{Zinnecker07} where the cross section for capturing a bypassing protostar is increased significantly due to the presence of an extended disc. In the following we examine the exact mechanism how the multiple protostellar systems are formed in our simulations.

\begin{figure}
 \centering
 \includegraphics[width=60mm]{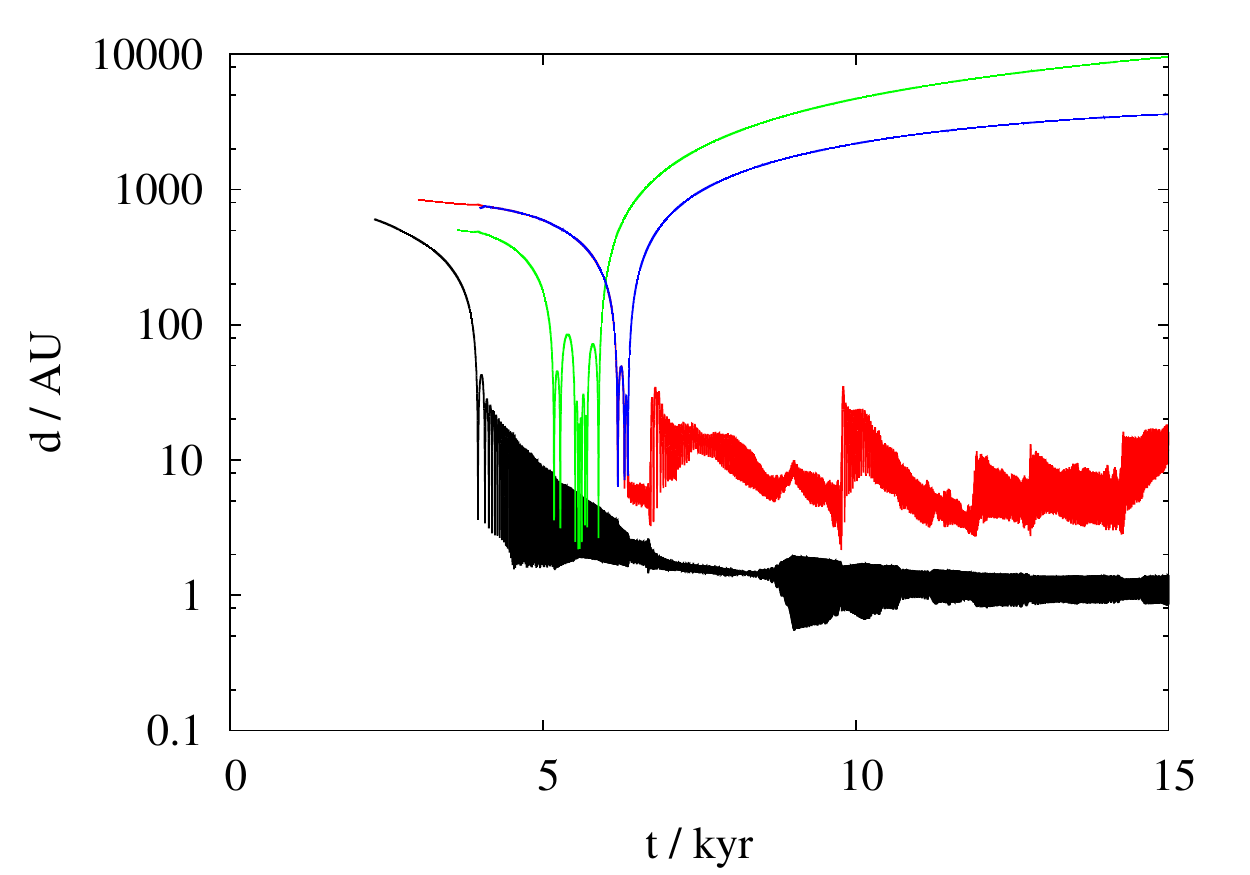}
 \caption{Distance of the different sink particles to the sink particle formed first (see Eq.~\ref{eq:distance}) for the simulation 2.6-Rot-M100.}
 \label{fig:dist}
\end{figure}

In Fig.\ref{fig:dist} we show the relative distance of each sink particle to the sink particle formed first for run 2.6-Rot-M100, i.e.
\begin{equation}
 d_\rmn{i}(\rmn{t}) = | \bmath{r}_\rmn{i}(\rmn{t}) - \bmath{r}_\rmn{1}(\rmn{t}) | \, ,
 \label{eq:distance}
\end{equation}
where $\bmath{r}_\rmn{i}(\rmn{t})$ is the position of the i-th sink particle at time $t$ and $ \bmath{r}_\rmn{1}(\rmn{t})$ that of the first sink particle.
As can be seen, by the time the sink particles form they are more than 100 AU away from the first sink particle and the centre of its associated disc. Hence, these sink particles do \textit{not} form by fragmentation of the already existing disc but clearly outside it during the collapse of distinct, gravitational unstable regions. Only during their further evolution do the secondarily formed sinks approach the first sink particle and are then captured in the disc. From Fig.~\ref{fig:dist} it can also be seen that two of the sinks get ejected from the disc due to close three-particle interactions and move away relatively quickly. These ejected sinks usually do not build up an associated disc around them as already pointed out in Section~\ref{sec:discs}.

\begin{figure}
 \centering
 \includegraphics[width=60mm]{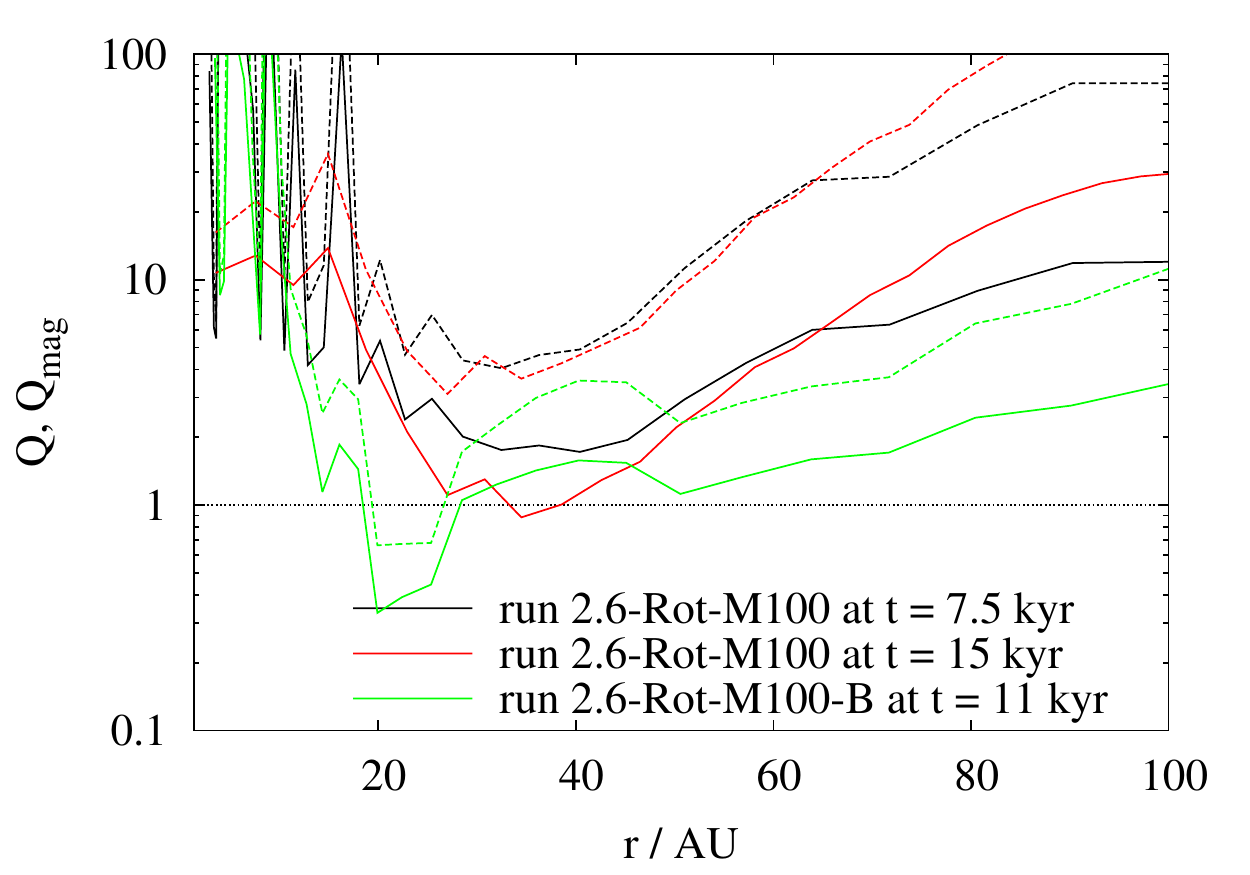}
 \caption{Toomre parameter $Q$ (solid lines) and magnetic Toomre parameter $Q_\rmn{mag}$ (dashed lines) for the runs 2.6-Rot-M100 and 2.6-Rot-M100-B at different times.}
 \label{fig:Toomre}
\end{figure}

At first sight it seems interesting that the sink particles in run 2.6-Rot-M100 do not form via disc fragmentation but rather due to the fragmentation of the entire core. For this reason we consider the Toomre parameter~\citep{Toomre64} describing the stability of the disc:
\begin{equation}
 Q = \frac{\kappa c_{\rmn{s}}}{\pi \Sigma G}
\end{equation}
with the epicyclic frequency $\kappa$, sound speed c$_{\rmn{s}}$, surface density $\Sigma$, and gravitational constant $G$. Gravitational instability sets in when $Q < 1$. As magnetic fields are present in the discs, a magnetic Toomre parameter~\citep{Kim01}
\begin{equation}
 Q_{\rmn{M}} = \frac{\kappa \left( c_{\rmn{s}}^2 + v_{\rmn{A}}^2 \right)^{1/2}}{\pi \Sigma G}
\end{equation}
can be defined as well, where $v_{\rmn{A}}$ is the Alfv\'enic velocity taking into account all components of the magnetic field. Since we know from Section~\ref{sec:discs} that the discs have almost Keplerian rotation velocities, we can easily replace the epicyclic frequency $\kappa$ in the above equations by the rotation frequency of the gas. In Fig.~\ref{fig:Toomre} we plot $Q$ and $Q_\rmn{mag}$ for the run 2.6-Rot-M100 at the same two times shown in Fig.~\ref{fig:diskM100}.
As can be seen, for both times $Q$ and $Q_\rmn{mag}$ are above the threshold of 1 indicating that the disc is indeed stable against gravitationally induced fragmentation. Only at r = 35 AU and t = 15 kyr does $Q$ drop below 1. However, since $Q_\rmn{mag}$ remains clearly above 1 and since no disc fragmentation is recognisable at that stage in the bottom right panel of Fig.~\ref{fig:diskM100}, this indicates that the magnetic field has a stabilising effect on the disc~\citep[see also][]{Hosking04,Machida05,Hennebelle08c,Duffin09,Seifried11}. We note that also for other times the results shown in Fig.~\ref{fig:Toomre} remain qualitatively unchanged. This supports our conclusion that the sink particles are formed outside the discs by core fragmentation as shown in Fig.~\ref{fig:dist}.

We emphasise that also for the runs 2.6-NoRot-M100, 2.6-Rot-M100-C, 2.6-Rot-M100-p2, 2.6-NoRot-M300 and 2.6-Rot-M1000, where more that one sink particle forms, fragmentation is not a consequence of a Toomre instability in the disc but due to the collapse of distinct gravitationally unstable regions. Hence, in these runs the occurrence of multiple protostellar systems is due to the capturing of bypassing protostars and not due to disc fragmentation. The only exemption is run 2.6-Rot-M100-B where in total 36 sink particles form. In this run after roughly 11 kyr disc fragmentation occurs around the most massive sink particle at a radius of about 25 AU. In this case both the classical and the magnetic Toomre parameter clearly drop below 1 shortly before the fragment forms as shown in Fig.~\ref{fig:Toomre} (green lines).

\section{Discussion}
\label{sec:discussion}

\subsection{Disc formation}

\subsubsection{Magnetic field misalignment}

In Section~\ref{sec:discs} we have shown that the first disc-like structures with noticeable rotational support build up around 5 kyr and are already well developed around 7.5 kyr (see Figs.~\ref{fig:diskM2} --~\ref{fig:velM300-1000}). Since the mass-to-flux ratio $\mu$, however, stays well below the critical value of 10 found in non-turbulent simulations over at least the first 10 kyr, no rotationally supported discs should be expected to form. Thus, the build-up of Keplerian discs from the very beginning on has to be attributed rather to the disordered magnetic field structure and the lack of a coherent rotation structure in the surroundings of the discs. The required amount of angular momentum is mainly carried by shear flows produced by the turbulent motions. Recently,~\citet{Hull12} reported a significant misalignment of outflows and magnetic fields in low-mass star forming regions. The misalignment was measured on scales of $\sim$ 1000 AU, thus on the same scale considered in Fig.~\ref{fig:muM2-100} in this work. As discussed in Section~\ref{sec:massflux}, our simulations suggest that the observed misalignment could also be a consequence of an unresolved disordered magnetic field structure. Moreover, since misaligned magnetic fields are simply a particular form of a disordered field structure, we suggest that the formation of Keplerian discs in the misaligned case~\citep{Hennebelle09,Ciardi10,Joos12} and in our work can be traced back to the same physical origin.

We note, however, that very recently~\citet{Li13} tried to confirm the results found by the authors mentioned before with a different numerical scheme. One striking result of their study is that misalignment does \textit{not} allow for Keplerian disc formation for $\mu < 5$, i.e. for typical observed values. Moreover, based on the work of~\citet{Joos12} and~\citet{Hull12},~\citet{Krumholz13} estimated the expected fraction of Keplerian discs to be as low as 10 -- 50 \%. In the presence of turbulence, however, we showed that in essentially any case -- even for $\mu < 5$ -- Keplerian discs can form.

\subsubsection{Magnetic flux loss}

Recently, \citet{Santos12} suggested that the loss of magnetic flux due to turbulent reconnection is responsible for the reduced magnetic braking efficiency. Magnetic flux loss alone, however, does not necessarily change the \textit{classical picture} of magnetic braking, i.e. a well-ordered magnetic field and coherent rotating as studied in numerous simulations neglecting turbulence. Hence, in this case the value of $\mu$ would be an appropriate quantity to assess the magnetic braking efficiency. However, in our simulations $\mu$ stays well below the critical value of $\sim$ 10 during the formation phase of the Keplerian discs (see Fig.~\ref{fig:muM2-100}) indicating that flux loss alone cannot account for the observed build-up of Keplerian discs. Thus, other mechanisms like the aforementioned disordered magnetic fields and turbulent motions in the close vicinity of the discs -- which are natural outcomes when using realistic initial conditions for star formation -- have to account for this.

In a second paper \citet{Santos12b} argue that, when considering scales of the size of the disc ($\sim$ 100 AU), $\mu$ will increase, a fact we also observe in our simulations. As mentioned in Section~\ref{sec:massflux}, the increase of $\mu$ could partly be due to accretion along the field lines. Moreover, we consider it as crucial to reduce the magnetic braking efficiency already on larger scales (by means of disordered magnetic fields but not weaker field strengths as indicated by the low value of $\mu$) since otherwise there would be no angular momentum left on the small scales (100 AU) to form the discs. This idea is also supported by the findings that for a well-ordered magnetic field the inwards transport of angular momentum is balanced by the (negative) magnetic torque at scales up to 500 -- 1000 AU~\citep{Seifried11}. Nevertheless, we suggest that the magnetic flux loss occurring subsequently in the evolution of the discs helps to maintain them in a rotationally supported state, in particular in the later phase (t $>$ 10 kyr).

Finally, we note that we did not cover the parameter space in a uniform way. Since we have a (at least) 5-dimensional parameter space spanning the strength of turbulence, magnetic fields and rotation as well as the core mass and the density profile, even a limited coverage of the parameter space would result in a considerable amount of different initial conditions. In combination with the very high resolution of 1.2 AU this leads to significant requirements of computational resources making a uniform coverage of the parameter space impossible. Nevertheless, even with the simulations presented here it becomes clear that turbulence-induced disc formation works for a wide range of molecular core masses and for different turbulence strengths. Furthermore, we showed that the formation of discs does not depend on the presence of an overall core rotation although the orientation of the discs does.

\subsection{Fragmentation properties}

The degree of fragmentation in most of the runs with 100 M$_{\sun}$ is relatively low ($\sim$ 5 sinks) given a highly gravitationally unstable core with supersonic turbulence. However,~\citet{Girichidis11} have shown that the fragmentation properties of a molecular cloud core with a density profile $\rho \propto r^{-1.5}$ as chosen in this work strongly depend on the realisation of the initial turbulence field. Indeed, in run 2.6-Rot-M100-B the core fragments much more heavily forming 36 sink particle in the first 15 kyr. The formation of Keplerian discs, however, is not affected by the different fragmentation behaviour.

However, since we do not include radiative feedback from the protostars in our simulations, in particular in the heavily fragmenting runs 2.6-Rot-M100-B, 2.6-NoRot-M300 and 2.6-Rot-M1000 we might significantly overestimate the number of fragments~\citep{Krumholz07,Krumholz10}. For the runs 2.6-NoRot-M100, 2.6-Rot-M100, 2.6-Rot-M100-C and 2.6-Rot-M100-p2, however, we find a similar number of fragments as in comparable simulations of~\citet{Krumholz07,Krumholz10} (5 -- 10 sinks), despite the fact that we did not include radiative feedback. This is most likely a consequence of the presence of strong magnetic fields. This was already stated by~\citet{Myers12}, who find that magnetic fields seem to suppress fragmentation in the diffuse core regions whereas radiative feedback mainly acts in the dense disc region.

However, also in our simulations without radiative feedback we find that disc fragmentation is not the main fragmentation mechanism in our simulations. In all runs the sink particles are formed via the gravitational collapse of distinct overdense condensations~\citep[][see also~\citealt{MacLow04} for an overview]{Klessen00,Heitsch01,Padoan04,Hennebelle08b}, only in run 2.6-Rot-M100-B a few sinks are formed via disc fragmentation. Comparing $Q$ and $Q_\rmn{mag}$ shows that the magnetic field can strongly contribute to the stability of the discs (Fig.~\ref{fig:Toomre}). We note that whereas in related work the reduced degree of fragmentation was often simply due to the suppression of Keplerian discs formation~\citep{Hosking04,Machida05,Hennebelle08c,Duffin09,Seifried11}, here it is indeed a consequence of rising the Toomre parameter above the critical value for stability. Hence, our result is complementary to the work of~\citet{Bate09}, \citet{Offner09,Offner10} and~\citet{Myers12}, who show that also radiative feedback suppresses the fragmentation of protostellar discs. For the accretion discs around high-mass stars, however,~\citet{Peters10,Peters11} have shown that ionization feedback and radiative heating is insufficient to strongly suppress fragmentation, at least outside of radii of $\sim 500\,$AU. Therefore it seems that the question of the importance of radiative feedback is not yet fully resolved.

The capture and inwards migration of secondarily formed sinks in the high-mass cases presents a path to form tight binaries (separation $\sim$ 1 AU). However, in run 2.6-NoRot-M100 and 2.6-Rot-M100 the binary forms after roughly 5 kyr (see Fig.~\ref{fig:dist}) with the lower-mass sink having a mass of $\sim$ 0.1 M$_{\sun}$. Hence, by that time an extended first core might still exist~\citep[e.g.][]{Masunaga98,Masunaga00}, which might have merged with the more massive sink to form a single protostar. We note that a more detailed study on the process of binary formation will be postponed to a subsequent study.

\subsection{Comparison to related work}
\label{sec:compare}

The initial conditions of run 2.6-NoRot-M300 were motivated by the recent work of~\citet{Myers12}. In their work the authors study the collapse of a molecular cloud core with similar properties under the additional influence of radiative feedback. Using a spatial resolution of 1.25 AU these authors find a 40 AU-sized Keplerian disc. With our run 2.6-NoRot-M300, which has a spatial resolution similar to that of~\citet{Myers12}, we can confirm this result finding even two Keplerian discs. The first disc in run 2.6-NoRot-M300 shows heavy fragmentation and temporarily gets completely disrupted but quickly re-establishes its Keplerian structure after each disruption event. Furthermore, we note that~\citet{Myers12} do not find a Keplerian disc when they reduce the spatial resolution by a factor of 8. We performed a resolution study for run 2.6-Rot-M100 with a four times and eight times coarser resolution, i.e. d$x$ = 4.7 and 9.4 AU, respectively. We find that with decreasing resolution the discs become smaller and progressively sub-Keplerian. Although there is still an indication of a rotationally supported structure in the run with a 8 times coarser resolution, the results of our resolution study strongly suggest that a spatial resolution of \textit{at least} a few AU -- depending on the actual numerical method used -- is needed to properly simulate the formation of Keplerian discs in a strongly magnetised environment.

Recently, the emergence of magnetic bubbles expanding radially outwards \textit{in the plane} of the disc have been observed in non-turbulent simulations~\citep[e.g.][]{Seifried11,Zhao11,Krasnopolsky12}. This feature, which occurs after a relatively short time (t $<$ 10 kyr), is a consequence of the ongoing mass accretion and -- due to the conditions of ideal MHD -- a corresponding magnetic flux accumulation. Once the magnetic pressure in the centre is strong enough to overcome the gravitationally drag, a low-density, highly magnetised bubble forms thus reducing the flux in the centre of the disc. Clearly, this bubble-like feature does not show up in our runs including turbulence. Since the accretion rates of the sink particles of 10$^{-5}$ M$_{\sun}$ yr$^{-1}$ and a few 10$^{-4}$ M$_{\sun}$ yr$^{-1}$ for the low-mass and high-mass runs, however, are comparable to the non-turbulent runs, one could naively expect the same amount of magnetic flux to accumulate in the disc centre eventually resulting in the appearance of a magnetic bubble. Since this is not the case, the magnetic field has to diffuse outwards before it is accreted onto the sink. This could be achieved by processes like magnetic reconnection and/or numerical diffusion in the already existing, rotationally supported disc. This agrees with the fact that $\mu$ increases in time (see top panel of Fig.~\ref{fig:muM2-100}) and for smaller radii~\citep[see also][]{Santos12b,Santos12}. A reason for an increased numerical diffusion of magnetic fields could be the lower infall velocities associated with Keplerian discs, which in turn give magnetic diffusion more time to operate. We note that the occurrence of numerical diffusion can also be motivated physically since in this regime non-ideal MHD effects like Ohmic dissipation or ambipolar diffusion are expected to come into play~\citep[e.g.][]{Nakano02,Mellon09,Duffin09,Dapp11}.

As pointed out in Section~\ref{sec:time}, the masses of the discs range from about 0.05 M$_{\sun}$ up to a few 0.1 M$_{\sun}$ and thus are a factor of about 10 below the sink particle masses. A simple density threshold criterion was used to define the disc. We note that recently~\citet{Joos12} suggested a more sophisticated criterion including also the velocity structure to determine the gas belonging to the disc. However, since both criteria can only give an approximate mass for the disc, we decided to stay with our simple criterion\footnote{We checked that a more sophisticated criterion similar to that of~\citet{Joos12} does not change the results significantly.}. For a more sophisticated comparison with observational data anyway synthetic observations produced with a radiation transfer code would we required.

Nevertheless, the disc masses in our runs agree reasonably well with a number of observations ranging from low-mass~\citep[e.g.][]{Jorgensen09,Enoch11} to high-mass protostellar objects~\citep[e.g.][]{Fuller01}. However, in particular for the latter case discs with significantly higher masses up to a few M$_{\sun}$ are observed, although these discs might correspond to structures without significant rotational support~\citep[e.g.][]{Shepherd01,Chini04,Fernandez11,Preibisch11}. Recently, Keplerian rotation in a Class 0 protostellar disc has been observed up to a radius of 90 AU~\citep{Tobin12} in accordance with our findings. In contrast,~\citet{Maury10} do not find Keplerian discs around Class 0 objects. Hence, to unambiguously answer the question at which stage Keplerian discs form, high-resolution observations e.g. with ALMA are required. Our work clearly shows that under realistic conditions, i.e. when including turbulent motions, early-type, rotationally supported protostellar discs can form in typical star forming regions.

\section{Conclusion} \label{sec:conclusion}

This study was aimed at testing the turbulence-induced disc formation mechanism reported in~\citet{Seifried12} for a wider range of initial conditions. We performed several simulations of collapsing molecular cloud cores with a wide range of masses (2.6 -- 1000 M$_{\sun}$) as well as different turbulence strengths ranging from sub- to supersonic. We showed that independently of the cores mass, the strength of turbulence, or the presence of global rotation in all cases Keplerian discs build up after a few kyr already.

The discs typically have a diameter of 50 -- 150 AU and masses of 0.05 up to a few 0.1 M$_{\sun}$. Interestingly, fragmentation mostly occurs in distinct overdense regions of the core rather than in the discs due to the Toomre instability. We showed that magnetic fields significantly contribute to the stability of the discs. Tight binaries can still form by disc capturing and inwards migration of secondarily formed protostars.

From the work on disc formation under the influence of magnetic fields done so far a common picture seems to arise: Due to the turbulent nature of star forming regions the classical picture of a well-ordered magnetic field structure and coherent rotational motions, used by~\citet{Mouschovias80} to assess the efficiency of magnetic braking, seems to break down. Under realistic conditions a disordered and/or misaligned magnetic field structure~\citep[see also][]{Hennebelle09,Ciardi10,Joos12,Joos13} as well as turbulent shear flows emerge, which cause the classical magnetic braking efficiency to drop and allow for the formation of Keplerian discs in essentially all cases considered here. The likelihood of disc formation in the turbulent case is therefore higher than that for purely misaligned fields in non-turbulent cores~\citep{Krumholz13,Li13}. During the formation of the discs the mass-to-flux ratio stays well below the critical value of 10, hence magnetic flux-loss alone cannot explain the formation of Keplerian discs. However, the subsequent increase of $\mu$ due to magnetic flux loss and mass accretion along magnetic field lines certainly helps to maintain a rotationally supported state during the further evolution of the discs.

\section*{Acknowledgements}

The authors like to thank the anonymous referee for the comments which helped to improve the paper. D.S. and R.B. acknowledge funding of Emmy-Noether grant 3706/1-1 by the DFG. D.S. furthermore gives thanks to IMPRS and HGSFP of the University of Heidelberg for financial support. R.E.P is supported by a Discovery grant from NSERC of Canada. R.S.K. acknowledges subsidies from the {\em Baden-W\"urttemberg-Stiftung} via contract research grant P-LS-SPII/18 and from the Deutsche Forschungsgemeinschaft (DFG) under grants no.\ KL 1358/11 and KL 1358/14 as well as via the Sonderforschungsbereich SFB 881 {\em The Milky Way System} (subprojects B1, B2, and B5). The simulations presented here were performed on Supermuc at the Leibniz Supercomputing Centre in Garching and on JUROPA at the Supercomputing Centre in J\"ulich. The FLASH code was developed partly by the DOE-supported Alliances Center for Astrophysical Thermonuclear Flashes (ASC) at the University of Chicago.

\label{lastpage}

\end{document}